\documentclass[journal]{IEEEtran}

\usepackage[usenames, dvipsnames]{color}
\usepackage{amsmath,amsfonts}
\usepackage[caption=false,font=normalsize,labelfont=sf,textfont=sf]{subfig}
\usepackage{textcomp}
\usepackage{stfloats}
\usepackage{verbatim}
\usepackage{array}
\usepackage{enumitem}
\usepackage{soul}
\usepackage{cite}
\usepackage{siunitx}
\usepackage{adjustbox}
\usepackage{hhline}
\usepackage{makecell}
\usepackage{caption}
\usepackage{float}
\usepackage{steinmetz}
\usepackage{tablefootnote}
\usepackage{scrextend}
\usepackage{multirow}
\usepackage{url, times, booktabs, tabularx, amssymb, bm, bbm}
\usepackage{algorithm, algorithmic}
\usepackage{bbding}
\usepackage{graphicx}
\usepackage[colorlinks=true, linkcolor=blue, urlcolor=blue, citecolor=blue]{hyperref}
\usepackage{ragged2e}

\def\BibTeX{{\rm B\kern-.05em{\sc i\kern-.025em b}\kern-.08em
    T\kern-.1667em\lower.7ex\hbox{E}\kern-.125emX}}
\usepackage{balance}

\begin{document}

\title{Selective-Memory Meta-Learning \\ with Environment Representations for \\ Sound Event Localization and Detection}

\author{Jinbo Hu,~\IEEEmembership{Student Member,~IEEE}, Yin Cao,~\IEEEmembership{Member,~IEEE}, Ming Wu,~\IEEEmembership{Member,~IEEE}, \\ Qiuqiang Kong,~\IEEEmembership{Member,~IEEE}, Feiran Yang,~\IEEEmembership{Member,~IEEE}, \\ Mark D. Plumbley,~\IEEEmembership{Fellow,~IEEE}, Jun Yang,~\IEEEmembership{Senior Member,~IEEE}
\thanks{
\scriptsize{

Jinbo Hu, Ming Wu, and Jun Yang are with the Key Laboratory of Noise and Vibration Research, Institute of Acoustics, Chinese Academy of Sciences, Beijing 100190, China (e-mail: hujinbo@mail.ioa.ac.cn; mingwu@mail.ioa.ac.cn; jyang@mail.ioa.ac.cn). Jinbo Hu and Jun Yang are also with the University of Chinese Academy of Sciences, Beijing 100049, China. \textit{(Corresponding author: Yin Cao; Jun Yang)}

Yin Cao is with the Department of Intelligent Science, Xi’an Jiaotong Liverpool University, Suzhou 215123, China (e-mail: yin.k.cao@gmail.com).

Qiuqiang Kong is with the Chinese University of Hong Kong, Hong Kong, China (e-mail: qqkong@ee.cuhk.edu.hk).

Feiran Yang is with the State Key Laboratory of Acoustics, Institute of Acoustics, Chinese Academy of Sciences, Beijing 100190, China (feiran@mail.ioa.ac.cn).

Mark D. Plumbley is with the Centre for Vision, Speech and Signal Processing, University of Surrey, Guildford GU2 7XH, U.K. (e-mail: m.plumbley@surrey.ac.uk).

This work was partially supported by the National Key Research and Development Project (NO. 2022YFB2602000), Grant "XJTLU RDF-22-01-084", and Engineering and Physical Sciences Research Council (EPSRC) Grant EP/T019751/1. For the purpose of open access, the authors have applied a Creative Commons Attribution (CC BY) license to any arising Author Accepted Manuscript version. This publication was supported by multiple datasets that are openly available at locations referenced in this paper.
}}}



\maketitle

\begin{abstract}

Environment shifts and conflicts present significant challenges for learning-based sound event localization and detection (SELD) methods. SELD systems, when trained in particular acoustic settings, often show restricted generalization capabilities for diverse acoustic environments. Furthermore, obtaining annotated samples for spatial sound events is notably costly. Deploying a SELD system in a new environment requires extensive time for re-training and fine-tuning. To overcome these challenges, we propose environment-adaptive Meta-SELD, designed for efficient adaptation to new environments using minimal data. Our method specifically utilizes computationally synthesized spatial data and employs Model-Agnostic Meta-Learning (MAML) on a pre-trained, environment-independent model. The method then utilizes fast adaptation to unseen real-world environments using limited samples from the respective environments. Inspired by the Learning-to-Forget approach, we introduce the concept of selective memory as a strategy for resolving conflicts across environments. This approach involves selectively memorizing target-environment-relevant information and adapting to the new environments through the selective attenuation of model parameters. In addition, we introduce environment representations to characterize different acoustic settings, enhancing the adaptability of our attenuation approach to various environments. We evaluate our proposed method on the development set of the Sony-TAu Realistic Spatial Soundscapes 2023 (STARSS23) dataset and computationally synthesized scenes. Experimental results demonstrate the superior performance of the proposed method compared to conventional supervised learning methods, particularly in localization.
\end{abstract}


\section{Introduction}
Sound event localization and detection (SELD) refers to detecting categories, presence, and spatial locations of different sound sources. SELD was first introduced in Task 3 of the Detection and Classification of Acoustics Scenes and Events (DCASE) 2019 Challenge \cite{dcase2019task3}. After three iterations of Task 3 of the DCASE Challenge \cite{dcase2019task3, dcase2020task3, dcase2021task3}, the types of data have transformed from computationally synthesized spatial recordings to real-scene recordings in 2022 and 2023 \cite{starss22,starss23}. Large-scale datasets of spatialized sound events were released for these challenges to be used for training and evaluating learning-based approaches.

\subsection{Learning-based SELD methods}
SELD can be regarded as a multi-task learning problem. Adavanne et al.\cite{Adavanne2018_JSTSP} proposed SELDnet for a joint task of sound event detection (SED) and regression-based direction-of-arrival (DOA) estimation. SELDnet cannot detect homogeneous overlap, which refers to overlapping sound events of the same type but from different locations. The Event-Independent Network V2 (EINV2), with a track-wise output format and permutation invariant training, was proposed to tackle the homogeneous overlap detection problem \cite{cao2020event,cao2021,hu2022track}. In contrast to the use of two outputs of SED and DOA in SELDnet and EINV2, the Activity-coupled Cartesian DOA (ACCDOA) approach merges the two subtasks into a single task\cite{shimada2021accdoa, multiaccdoa}, where the Cartesian DOA vectors also contain the activity information of sound events. 

However, the performance of learning-based methods is usually degraded when the training set and test set are mismatched. The training set cannot cover all actual instances from different acoustic environments.

\subsection{Environment shifts and conflicts}

The change in the data distribution between a training set and a test set is known as the \textit{domain shift} problem \cite{DA-survey, DA-survey-tpami}. In the practical deployment of SELD systems, the differences among environments could potentially be very significant. Unseen complex acoustic environments could result in a decline in system performance, due to the distribution change in acoustic properties, such as varying degrees of echo and reverberation, diverse types of ambient noise, and directional interference. The change in the distribution of acoustic properties among environments is referred to as the environment shift. The issue is particularly salient when the system encounters acoustic properties it has not been exposed to during training. On the other hand, different acoustic environments may present conflicts. Optimal solutions for diverse acoustic environments may display considerable variation, e.g., indoor environments and outdoor environments. A single training configuration cannot encompass all types of environments.

\begin{figure}[t]
    \centerline{\includegraphics[width=0.95\linewidth]{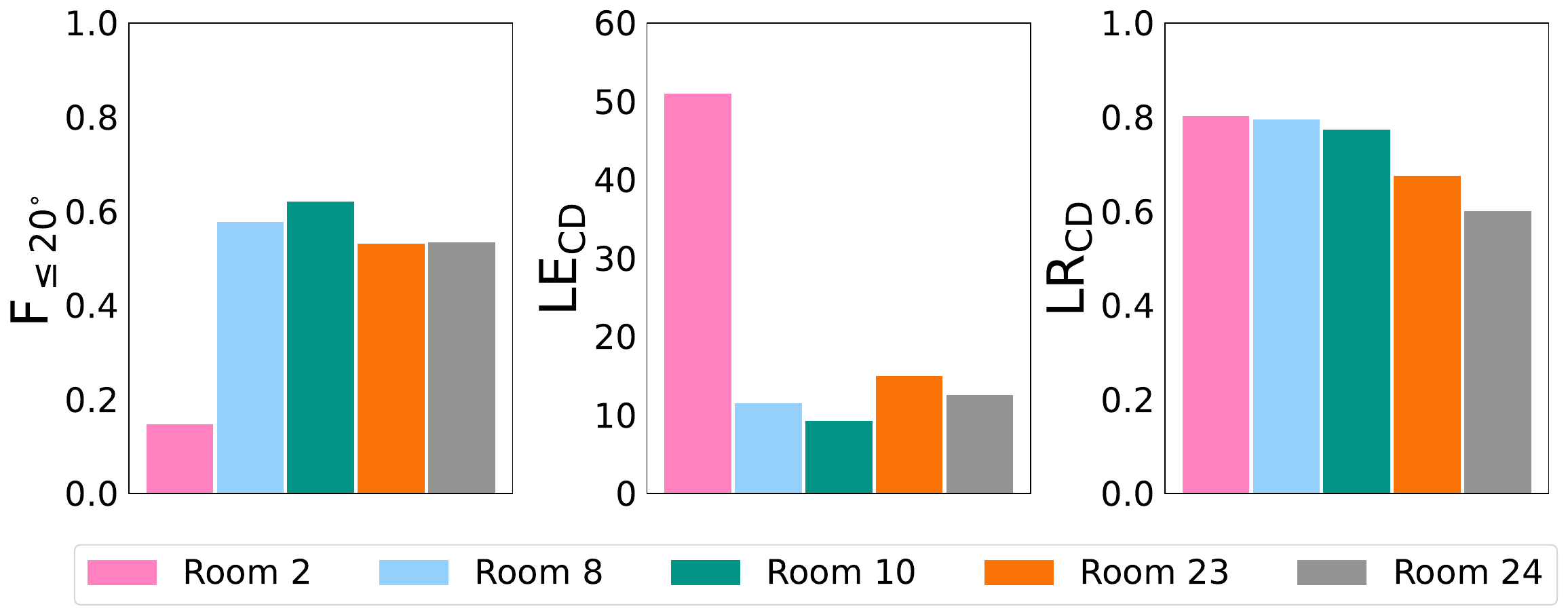}}
    \caption{Room-wise metric scores of our previous system \cite{hu22dw} submitted to Task 3 of the DCASE 2022 Challenge on the STARSS22 validation set. The description of each metric is expounded in Section \ref{sec: metric}. The system obtained the second rank in the team ranking.}
    \label{fig:starss22dev}
\vspace{-2mm}
\end{figure}

Fig. \ref{fig:starss22dev} illustrates the results of our previous system \cite{hu22dw} submitted to Task 3 of the DCASE 2022 Challenge. STARSS22 \cite{starss22} is a dataset of spatial recordings of real scenes with spatiotemporal annotations of sound events. There are no duplicated recording environments between the training and test sets in STARSS22. The system was evaluated on the STARSS22 dataset and obtained second in the team ranking. However, we found unsatisfactory generalization performance for the Room 2 recordings in the \textit{dev-test-tau} set of STARSS22 \cite{hu22dw} compared with other rooms. Experimental results show that class-dependent localization error $\mathrm{LE}_\mathrm{CD}$ is much higher, and location-dependent F-score $\mathrm{F}_{\leq 20^\circ}$ is much lower in Room 2, but class-dependent localization recall $\mathrm{LR}_\mathrm{CD}$ is high. This suggests that our system may have weak localizing performance in Room 2 due to the environment shift or conflict.

\subsection{Data acquisition}

One of the most effective methods for solving environment shifts and conflicts is acquiring as much data as possible \cite{DA-survey, DA-survey-tpami, grumiaux2022survey}. However, manually collecting and annotating spatial sound event recordings is highly cost-intensive. For instance, the STARSS22 dataset \cite{starss22} employed sophisticated setups involving a 32-channel spherical microphone array for recording, wireless microphones for manually annotating types, onset, and offset of sound events, a motion capture (mocap) system for extracting the tracked data, and a $360^\circ$ camera for validating those annotations.

Another practical approach for acquiring data is computationally synthesizing spatial sound event samples. Spatial sound event recordings are simulated by convolving dry sound event samples with spatial room impulse responses (SRIRs). The multi-channel simulation (MCS) framework \cite{wang2023four} and the impulse response simulation (IRS) framework \cite{koyama2022spatial} are both designed to simulate multi-channel recordings, emphasizing augmentation from original data without the reliance on external datasets. MCS and IRS involve the convolution of enhanced source signals with extracted covariance matrices and computationally simulated SRIRs, respectively. Noteworthy among existing external datasets used for synthesizing spatial sound events include FSD50K \cite{fonseca2021fsd50k}, AudioSet \cite{gemmeke2017audio}, and TAU-SRIR DB \cite{srir-db}. FSD50K and AudioSet are large-scale sound event datasets, while TAU-SRIR DB is a real-recorded SRIR database tailored for the DCASE Challenge. Given the insufficiency of publicly accessible SELD data, particularly for specific microphone array types, e.g., the number of microphones and geometry of the microphone array, the role of the SRIR simulation technique becomes indispensable in data synthesis. Numerous RIR simulation methodologies rely on geometric approaches \cite{allen1979image, pygsound}, where the propagation of sound waves is modeled and manipulated in the form of a ray. These techniques simulate the reflections and reverberation of sound waves. Software packages like Pyroomacoustics \cite{pyroomacoustics}, gpuRIR \cite{gpurir}, and SMIR-Generator \cite{smir-generator} are representative tools for geometric-based SRIR simulations.

\subsection{Meta learning}
Models trained on synthetic datasets typically demonstrate a degree of generalization ability but could limit the robustness of the network on real-world data due to the environment shift \cite{grumiaux2022survey}. One line of research is to train the model with realistic signals by transfer learning \cite{survey-transfer-learning}, pre-training a model with large-scale synthetic datasets, and then fine-tuning the model with limited real-recorded datasets \cite{grumiaux2022survey}. 

In circumstances when environmental disparities occur between training sets and test sets and collected sample availability is limited, few-shot learning (FSL) \cite{wang2020fslsurvey} can offer a solution for adaptation to realistic and specific environments. 

The conventional supervised learning method refers to training a model on a labeled dataset and then applying the trained model to predict unseen data. The core issue of conventional supervised learning training methodologies in the case of limited samples is that the empirical risk minimizer is unreliable \cite{wang2020fslsurvey}. Nevertheless, by incorporating prior knowledge, FSL can effectively generalize to the specific task, even with only a few samples \cite{wang2020fslsurvey}. Meta-learning, which facilitates FSL, learns a general-purpose learning algorithm that generalizes across tasks and ideally enables each new task to be learned well from the task-distribution view \cite{meta-survey}. Meta-learning has advanced FSL significantly in computer vision \cite{closerlook, mattersvison}. 

One of the most successful meta-learning algorithms is model-agnostic meta-learning (MAML) \cite{maml}. MAML formulates prior knowledge as commonly initialized parameters across tasks and then exploits a few samples of the target task to adapt that task quickly. Due to its model-agnostic nature, MAML is compatible with any model trained with gradient descent, making it applicable to various learning problems, including classification, regression, and reinforcement learning. In audio signal processing, the meta-learning method has recently attracted interest in solving FSL problems. Based on MAML, several audio-related researchers have investigated building systems to adapt their specific tasks rapidly with only a few corresponding samples \cite{meta-tts,meta-localization,meta-tdsv}. To the best of our knowledge, the few-shot environment adaptation problems based on meta-learning have not been thoroughly studied. 

MAML can be designed to employ a set of trained initial parameters and a few samples from a specific environment to cope with the environment shift problem and train multiple models for each specific environment to somewhat mitigate the environment conflict issue. However, forcibly sharing the initial parameters can still lead to some conflicts and compromises among tasks \cite{l2f}. Multimodal MAML (MMAML) \cite{mmaml} focuses on task-dependent initial parameters and tries to learn task embeddings and transform the initial parameters with affine parameters. Compared with Multimodal MAML, Learning-to-Forget (L2F) \cite{l2f} proposes layer-wise attenuation on the compromised initial parameters for each task to reduce its influence. 

\subsection{Our contributions}

\begin{figure*}[t!]
  \centering
  \centerline{\includegraphics[width=0.95\textwidth]{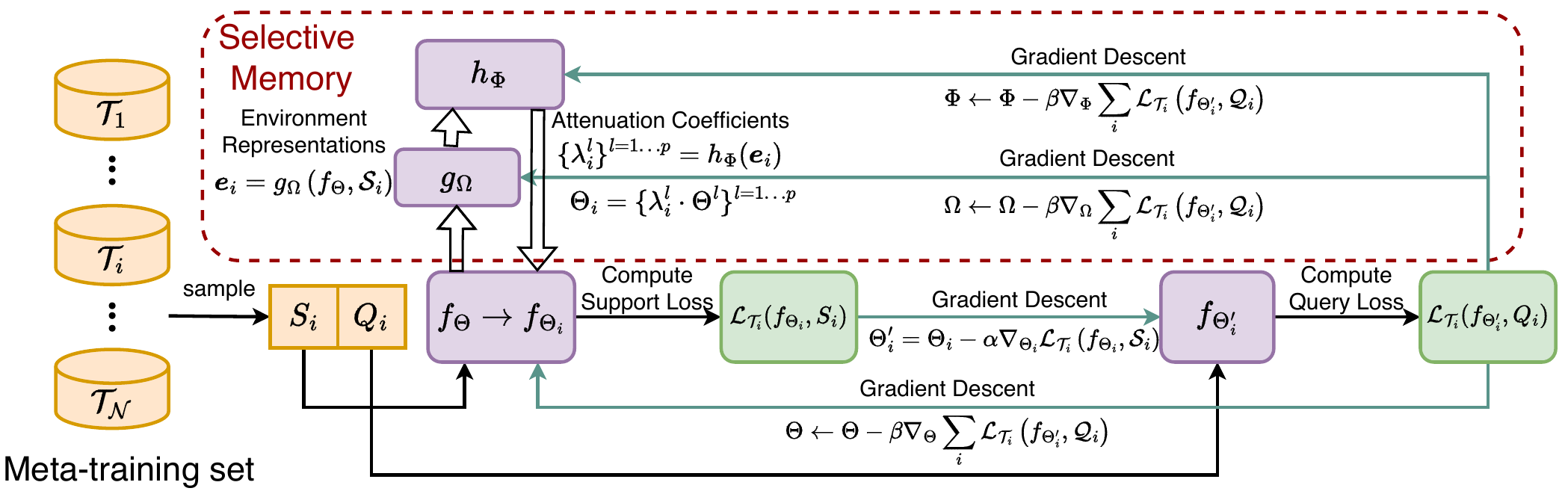}}
  \caption{A diagram of the meta-training procedure for our proposed environment-adaptive Meta-SELD. For simplicity, we only consider one gradient update for the inner loop of the training procedure. $\mathcal{N}$ indicates the number of tasks in the meta-training set. $f_\Theta$, $g_\Omega$, and $h_\Phi$ represent the backbone, environment extractor, and attenuation network, respectively. $\Theta^l$, where $l=1\dots p$, denotes the $l$-th layer of the total $p$-layer backbone $f_\Theta$.}
  \label{fig: maml+l2f-diagram}
\vspace{-2mm}
\end{figure*}

In this work, we extend our previous work Meta-SELD \cite{meta-seld} to environment-adaptive Meta-SELD, investigating an adaptation to the environment shift problem using meta-learning-based few-shot methods. Drawing inspiration from L2F \cite{l2f}, we propose to selectively memorize components relevant to the target environment and to learn the target environment by using environment representations. Fig. \ref{fig: maml+l2f-diagram} presents a flow diagram of environment-adaptive Meta-SELD. In contrast to \cite{zsl-seld}, which tackles unseen categories of sound events problem in SELD, our proposed Meta-SELD focuses on adaptation to unknown environments.

\begin{figure}[t]
  \centerline{\includegraphics[width=0.98\linewidth]{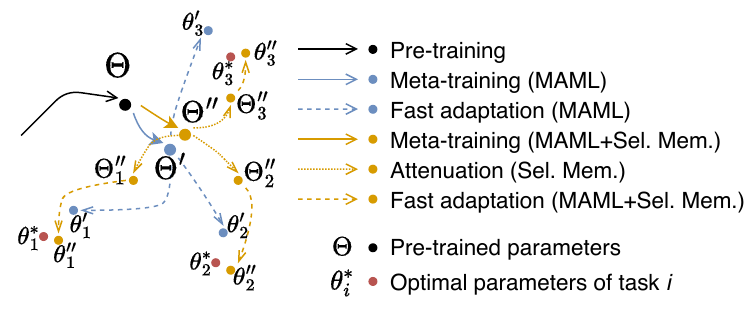}}
  \caption{An illustration of Meta-SELD++ with and without selective memory. Selective memory is proposed to tackle environment conflicts. It adds an additional step for attenuation of initial parameters before fast adaptation to the task $i$. The selective memory method provides a better solution (closer to the optimal solution).}
  \label{fig: maml+l2f-opt}
\vspace{-2mm}
\end{figure}

The method of fast adaptation to environments is mainly based on Model-Agnostic Meta-Learning (MAML) \cite{maml} and an environment-independent (EI) model. The EI model is pre-trained on a computationally synthesized dataset, encompassing a wide range of acoustic environments. We then apply MAML to the pre-trained EI model to create a meta-EI model, which enables fast adaptation to an unseen environment using a few samples recorded in the target environment. In addition, forcibly sharing common initial parameters across environments allows the environment conflict issue to remain. Inspired by Learning-to-Forget (L2F) \cite{l2f}, we adopt an attenuation network and propose environment representations to selectively memorize target-environment-relevant components and selectively forget contradicting information. An illustration of our proposed environment-adaptive Meta-SELD is depicted in Fig. \ref{fig: maml+l2f-opt}. We evaluated the proposed method on the development set of the Sony-TAu Realistic Spatial Soundscapes 2023 (STARSS23) \cite{starss23} dataset and on computationally synthesized scenes. Experimental results demonstrated the effectiveness of environment-adaptive Meta-SELD.

Our main contributions can be summarized as follows:

1) Pre-training an environment-independent (EI) model on computationally synthesized datasets to contain as many acoustic properties as possible.

2) Investigating an environment adaptation approach based on Model-Agnostic Meta-Learning (MAML) and the pre-trained EI model.

3) Introducing a solution to selectively memorize prior information relevant to the target environment to mitigate environment conflicts.

4) Proposing a technique to extract environment representations for selective memory and designing comprehensive experiments to display environment representations.

\section{Fast adaptation to the environment}

\subsection{Pre-trained environment-independent models}
We train a simple convolutional recurrent neural network (CRNN) on the datasets synthesized using computationally generated SRIRs. Source code about data synthesis of spatial sound events has been released\footnote{https://github.com/Jinbo-Hu/SELD-Data-Generator}. This model is not trained on samples in specific environments and is consequently termed the environment-independent (EI) model.

\subsubsection{Data synthesis}
The spatial sound event recordings are simulated by convolving monophonic sound event samples with SRIRs. The sound event samples are selected from FSD50K and AudioSet, based on the similarity of the labels in those datasets to target classes in STARSS23. To acquire high-quality clips, these clips are subsequently filtered by the pre-trained CNN14 \cite{kong2020panns} model based on the inference probability. The SRIRs are computationally generated using geometric-based methods \cite{allen1979image, pygsound}. The computational generation method for SRIRs consists of two steps: microphone-array RIRs simulation and Ambisonics format converter \cite{rafaely2015fundamentals, archontisPhD, politis2017comparing}.

We use the image source method \cite{allen1979image} to generate microphone-array RIRs with four channels. This method replaces reflection on walls with virtual sources playing the same sound as the source and builds an RIR from the corresponding delays and attenuations. The procedure of the Ambisonics format converter is described in the Appendix.

\subsubsection{Network architecture}
\begin{figure}[t]
  \centering
  \centerline{\includegraphics[width=0.9\linewidth]{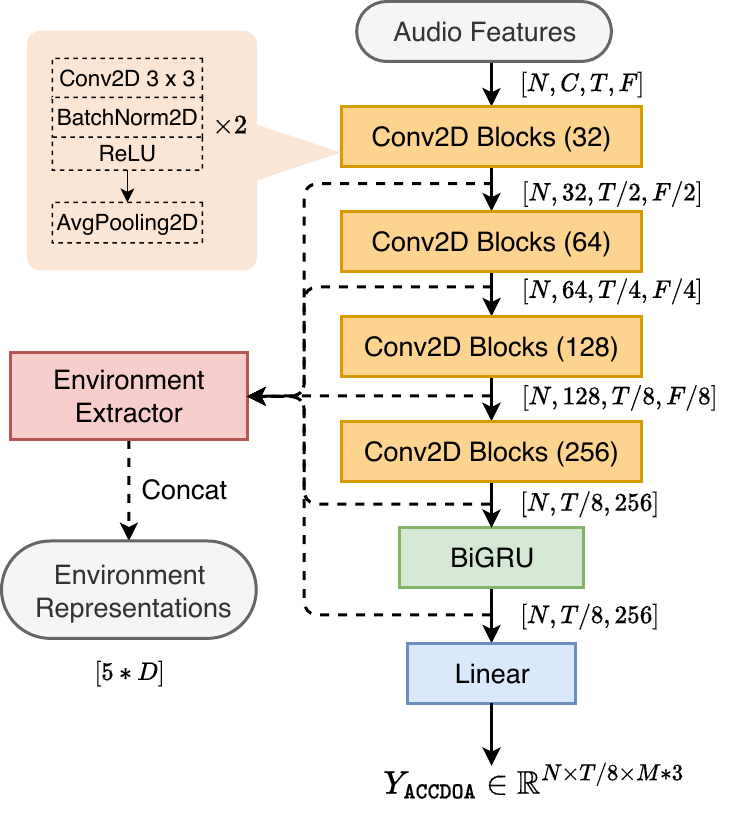}}
  \caption{The network architecture of the SELD network with the sub-network of environment representation extraction. The environment representations are extracted from output feature maps of each layer of the backbone.}
  \label{fig: maml+l2f-crnn}
\end{figure}

Without loss of generality, in this study, we adopt a simple CRNN as our backbone for our following experiments. The CRNN is similar to the baseline of Task 3 of the DCASE 2022 Challenge\cite{starss22} but with an ACCDOA representation \cite{shimada2021accdoa}. As shown in Fig. \ref{fig: maml+l2f-crnn}, the backbone has four convolution blocks followed by a one-layer bidirectional gated recurrent unit (BiGRU). The network takes $C$-channel $T$-frame $F$-mel-bin spectrograms, the concatenation of log-mel spectrograms and intensity vectors as input, and predicts active sound events of $M$ classes with corresponding Cartesian DOA vectors for each time stamp. 

\subsection{Meta-SELD}
Given a model represented by a parameterized function $f_\Theta$ with parameters $\Theta$, MAML \cite{maml} learns the initial parameters $\Theta_{0}$ from general tasks $\mathcal{T}_i$ sampled from the meta-training set $\mathcal{D}_\mathtt{train}$. The initial parameters $\Theta_{0}$ are sensitive to task-specific fine-tuning \cite{maml} and expected to perform well on unseen tasks from the meta-test set $\mathcal{D}_\mathtt{test}$ after a few parameter updates with a few task-specific samples. The task in meta-learning refers to a specific learning problem. Each task $\mathcal{T}_i$ consists of a support set $\mathcal{S}_i$ of $K$ samples and a query set $\mathcal{Q}_i$ of $Q$ samples, analogy to the training set and test set, respectively. A new task is expected to be quickly learned with $K$ samples, known as $K$-shot learning. The loss function of MAML is
\begin{equation}
\label{eq: MAML loss}
\mathcal{L} = \sum_{\mathcal{T}_i\sim P(\mathcal{T})} \mathcal{L}_{\mathcal{T}_i}(f_{\Theta})
\end{equation}
where $P(\mathcal{T})$, sampled from $\mathcal{D}_\mathtt{train}$, is a distribution over tasks that we want our model to adapt to, and $\mathcal{L}_{\mathcal{T}_i}$ indicates the task-specific loss. In essence, the loss function $\mathcal{L}$ evaluates the ability for adaptation of a model. It measures the task-specific loss for a batch of tasks after task-specific parameter updates based on a common set of initial parameters $\Theta$. In contrast to conventional supervised learning methods, the objective of which is to find optimal parameters to minimize the loss function across all training samples, MAML tries to find common generalized initial parameters across tasks and then updates the initial parameters after several iterations of training on limited data of new tasks.

There are two groups of parameters in the MAML algorithm: meta parameters and the task-specific adaptation parameters. In the meta-training phase, MAML starts with randomly initialized meta parameters $\Theta$ and then adapts to a new specific task $\mathcal{T}_i$ with several update iterations using the support set $\mathcal{S}_i$. We initialize the adaptation parameters of the task $\mathcal{T}_i$ with meta parameters $\Theta_i^\prime=\Theta$, and update $\Theta_i^\prime$ by
\begin{equation}
\label{eq: innerloop}
{\Theta}_i^\prime \gets {\Theta}_i^\prime-\alpha \nabla_{{\Theta}_i^\prime} \mathcal{L}_{\mathcal{T}_i}\left(f_{{\Theta}_i^\prime}, \mathcal{S}_i\right)
\end{equation}
where $\alpha$ is the adaptation learning rate. After computing task-specific loss on $\mathcal{Q}_i$ with $f_{{\Theta}_i^\prime}$ across a batch of tasks, the meta parameters are updated as follows:
\begin{equation}
\Theta\gets\Theta-\beta \nabla_{\Theta} \sum_i \mathcal{L}_{\mathcal{T}_i}\left(f_{{\Theta}_i^\prime}, \mathcal{Q}_i\right)
\end{equation}
where $\beta$ is the meta step size. Mathematically, the updates of meta parameters involve a gradient through a gradient, which necessitates an additional backward pass through $f$ to compute the Hessian-vector product \cite{maml}. For reducing computation and memory burdens, we omit this backward pass and instead employ a first-order approximation, which also achieves comparable performance in contrast to the second-order derivative \cite{maml}. After accumulating $\mathcal{L}_{\mathcal{T}_i}$ for several tasks, $\Theta$ is updated by gradient descent and will be used as the initial parameters for the subsequent loops of meta-training steps.

In the meta-testing phase, a specific unseen task $\mathcal{T}_j^\mathtt{test}$ created using the meta-test set $\mathcal{D}_\mathtt{test}$ is used. $\mathcal{T}_j^\mathtt{test}$ consists of a labeled support set $\mathcal{S}_j^\mathtt{test}$ of $K$ samples, and an unlabeled query set $\mathcal{Q}_j^\mathtt{test}$ of $Q$ samples. We update the model, initialized by well-trained parameter $\Theta$ in the meta-training phase, on $\mathcal{S}_j^\mathtt{test}$ to get adaptation parameters ${\Theta_j}^\prime$ using the update procedure of Eq. \ref{eq: innerloop}. The adaptation performance is evaluated on $\mathcal{Q}_j^\mathtt{test}$ with $f_{\Theta_j^\prime}$.

We aim to adapt to an unseen environment with $K$ samples ($K$-shot learning). The objective of MAML is to find optimal initial parameters across several tasks, so we need to construct a set of tasks from the meta-training set $\mathcal{D}_\mathtt{train}$. $\mathcal{D}_\mathtt{train}$ is split into several tasks according to the different recording environments. Audio clips recorded in different environments belong to different tasks. We first sample a batch of environments and then sample $K+Q$ clips in each environment, where $K$ clips from the support set $\mathcal{S}_i$ and $Q$ clips from the query set $\mathcal{Q}_i$. 

\begin{figure}[t]
    \centering    
    \includegraphics[width=0.95\linewidth]{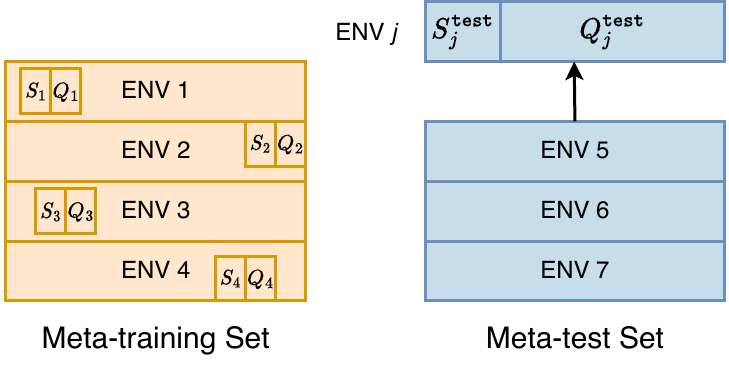}
    \caption{The data division for the meta-training and the meta-test sets according to recording environments.}
    \label{fig: dataset split}
\end{figure}

The meta-learning processes of SELD for testing and training are slightly different in the data division. Fig. \ref{fig: dataset split} shows the division of the meta-training and the meta-test sets. Similar to the meta-training set $\mathcal{D}_\mathtt{train}$, the meta-test set $\mathcal{D}_\mathtt{test}$ is partitioned based on the recording environments of each audio clip. For clips of each environment, we also chose $K$ clips for meta-test support set $\mathcal{S}_j^\mathtt{test}$, and all remaining clips for meta-test query set $\mathcal{Q}_j^\mathtt{test}$. After $N$ iterations of update on $\mathcal{S}_j^\mathtt{test}$, the meta parameters $\Theta$ are updated to ${\Theta_j^\prime}$. The final performance is evaluated on $\mathcal{Q}_j^\mathtt{test}$ with $f_{\Theta_j^\prime}$.

\subsection{Meta-SELD++}
Instead of randomly initializing the meta parameters, we leverage the power of the pre-trained environment-independent (EI) model to initialize the meta parameters of MAML. In other words, we may utilize a well-trained EI model to find better initial parameters and accelerate convergence for training the meta-EI model. 

Specifically, during meta-training, we initialize the meta parameters with parameters of the pre-trained EI model and then create a meta-EI model trained on synthetic datasets using collected SRIRs based on MAML. We denote this method as Meta-SELD++, whose meta parameters are initialized with pre-trained parameters. The training procedure will be expounded in Section \ref{sec: datasets}.

\section{Environment-Adaptive Meta-SELD}
\subsection{Selective memory}
According to Eq. \ref{eq: MAML loss}, MAML gives each task equal weight and tries to find optimal initial parameters across tasks in an average sense so that it may perform better or worse on a specific task than the conventional supervised learning method. There may be conflicts when optimizing across a batch of environments \cite{meta-seld}. Inspired by L2F \cite{l2f}, which argues that forcibly sharing a common initialization in MAML induces conflicts and thus leads to the compromised location of the initialization, we selectively memorize parts relevant to the target environment. More comprehensively, we employ an environment-dependent layer-wise attenuation on the initialization, thereby dynamically controlling the influence of prior knowledge for each environment. The attenuation is generated by an attenuation network $h_\Phi$ with random initialization. Fig. \ref{fig: maml+l2f-opt} illustrates the optimization procedure of selective memory Meta-SELD. Compared with Meta-SELD++, selective memory adds a step to attenuate the common initial parameters for target environments before fast adaptation to corresponding environments.

One general information input to $h_\Phi$ is gradients. At the beginning of the inner loop of MAML, task-specific gradients $\nabla_{\Theta} \mathcal{L}_{\mathcal{T}_i}\left(f_{\Theta}, \mathcal{S}_i\right)$ on the support sets $\mathcal{S}_i$ of the $i$-th environment are computed to generate layer-wise attenuation coefficients:
\begin{equation}
\boldsymbol{\lambda_i}=h_\Phi\left(\nabla_{\Theta} \mathcal{L}_{\mathcal{T}_i}\left(f_{\Theta}, \mathcal{S}_i\right)\right)
\end{equation}
where $h_\Phi$ is an MLP network with parameters $\Phi$, consisting of a ReLU non-linearity sandwiched between two linear layers with the hidden size of 1024 by default, and a sigmoid in the final layer to facilitate attenuation. The layer-wise attenuation coefficients $\boldsymbol{\lambda_i}=\{\lambda_i^l\}^{l=1\dots p}$ act on each layer of $\Theta=\{\Theta^l\}^{l=1\dots p}$:
\begin{equation}
\begin{split}
    \Theta_i &= \boldsymbol{\lambda_i} \odot \Theta \\
    &= \{ \lambda_i^l \cdot \Theta^l\}^{l=1 \dots p}
\end{split}
\end{equation}
where $l$, $p$ and $\odot$ indicate the layer index, the number of layers of the backbone $f_\Theta$ and Hadamard product, respectively. Note that $\lambda_i^l$ is a single value applied to all parameters within the $l$-th layer.

Task-specific gradients generate attenuation coefficients insensitive to various environments and hence seem to be ineffective information to make attenuation environment-adaptive, which will be presented and analyzed in Section \ref{sec: exp. sel_mem}. We adopt novel representations relevant to environments and expect that these representations can effectively capture and represent the acoustic properties of various environments. These representations are acquired through an unsupervised learning approach, enabling the selective memorization of target-environment-dependent prior knowledge to some extent. We denote selective memory Meta-SELD with environment representations as \textit{environment-adaptive Meta-SELD}.

\subsection{Environment representations}

\begin{figure}[t]
    \centering
    \includegraphics[width=0.98\columnwidth]{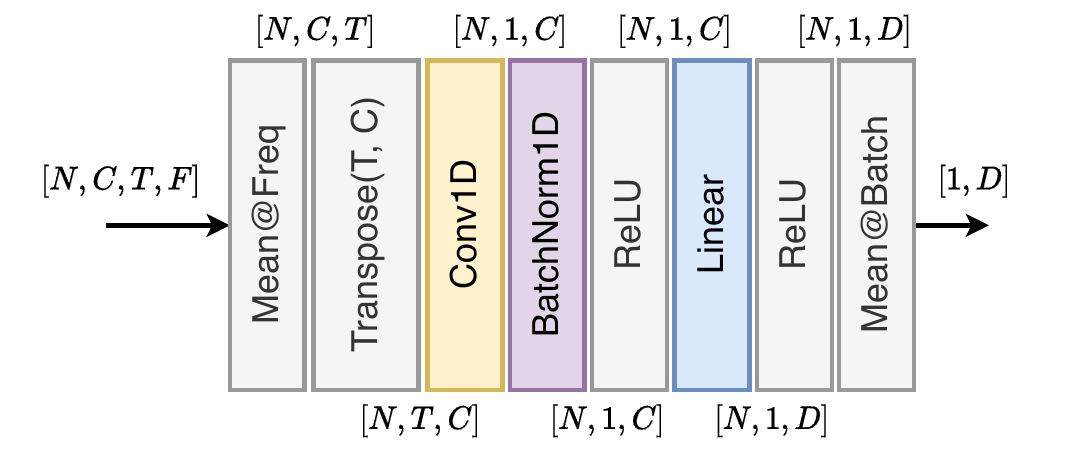}
    \caption{A detailed implementation of the environment extractor.}
    \label{fig: emb-extract}
\end{figure}

Fig. \ref{fig: maml+l2f-crnn} depicts the method of extracting environment representations. Specifically, we employ the environment extractor $g_\Omega$  with random initialization to extract feature embeddings from output feature maps of each backbone layer. Subsequently, feature embeddings from each layer are concatenated to environment representations. The detail of the environment extractor is shown in Fig. \ref{fig: emb-extract}. The feature embedding dimension $D=2048$ is used according to demonstrated performance. The output feature maps of each layer are averaged and weighted-averaged in the batch and time dimension. We conjecture that these operators mitigate the influence of acoustic events on the feature embeddings while preserving the environmental information. Fig. \ref{fig: maml+l2f-diagram} and Algorithm \ref{alg: MAML+L2F} illustrate and summarize the overall training procedure of environment-adaptive Meta-SELD. Step 6 in Algorithm \ref{alg: MAML+L2F} represents the extractor process of environment representations, Step 11 is the inner-loop update for adaptation parameters, and Step 15 is the outer-loop update for meta parameters. 

Our emphasis lies in evaluating the performance of Meta-SELD in unfamiliar environments. We hypothesize that audio recordings captured in diverse spatial locations within a given environment exhibit more similar acoustic properties when compared to recordings from different environments, excluding extreme cases where the speaker or microphone is close to reflections or each other. Furthermore, these acoustic properties generally remain consistent across recording moments and sound events. To test our hypothesis, we display environment representations via the similarity map and t-SNE \cite{t-sne} in Section \ref{sec: exp. env_rep}.

\begin{algorithm}[t]
\caption{Meta-training of proposed environment-adaptive Meta-SELD}
\label{alg: MAML+L2F}
\begin{algorithmic}[1]
    \REQUIRE Distribution over all environments $P(\mathcal{T})$, adaptation step size $\alpha$, meta step size $\beta$
    \REQUIRE Parameters $\Theta=\{\Theta^l\}^{l=1\dots p}$ from pre-trained EI model, where $l$ is the layer index and $p$ is the number of layers of the backbone
    \STATE Initialize meta parameters of backbone $f_\Theta$ with the pre-trained parameters $\Theta$. Randomly initialize environment extractor $g_\Omega$ and attenuation network $h_\Phi$ in Fig. \ref{fig: maml+l2f-diagram}
    \FOR{$\text{epoch}_{it}=1,2,\ldots,N_\text{epochs}$}
        \STATE Sample a permutation of environments without replacement encompassing all environments $\mathcal{T}_i$ from $P(\mathcal{T})$
        \FOR {each environment $\mathcal{T}_i$}
            \STATE Sample disjoint examples $(\mathcal{S}_i, \mathcal{Q}_i)$ from $\mathcal{T}_i$
            \STATE Compute representations of the $i$-th environment: $\boldsymbol{e_i}=g_\Omega(f_\Theta, \mathcal{S}_i)$
            \STATE Compute attenuation coefficients $\boldsymbol{\lambda_i}=\{\lambda_i^l\}^{l=1\dots p}$ for each layer: $\boldsymbol{\lambda_i}=h_\Phi(\boldsymbol{e_i})$
            \STATE Compute attenuated initial parameters: $\Theta_i^l=\lambda_i^l \cdot \Theta^l$
            \STATE Let $\Theta_i^\prime = \{\Theta_i^l\}^{l=1\dots p}$
            \FOR{gradient descent step $t:=0$ \textbf{to} $N-1$}
                \STATE Perform gradient descent to update adaptation parameters: $\Theta_{i}^\prime \gets \Theta_{i}^\prime - \alpha \nabla_{\Theta_i^\prime}\mathcal{L}_{\mathcal{T}_i}\left(\Theta_{i}^\prime, \mathcal{S}_i\right)$
            \ENDFOR
            \STATE Compute $\mathcal{L}_{\mathcal{T}_i}(f_{\Theta_i^\prime}, \mathcal{Q}_i)$ by evaluating $\mathcal{L}_{\mathcal{T}_i}$ w.r.t. $\mathcal{Q}_i$ and $\Theta_i^\prime$ of the $i$-th environment
        \ENDFOR
        \STATE Perform gradient descent to update meta parameters:
        $\Theta \gets \Theta - \beta\nabla_{\Theta}\sum_i\mathcal{L}_{\mathcal{T}_i}\left(f_{\Theta_i^\prime}, \mathcal{Q}_i\right)$ \\
        $\Omega \gets \Omega - \beta\nabla_{\Omega}\sum_i\mathcal{L}_{\mathcal{T}_i}\left(f_{\Theta_i^\prime}, \mathcal{Q}_i\right)$ \\
        $\Phi \gets \Phi - \beta\nabla_{\Phi}\sum_i\mathcal{L}_{\mathcal{T}_i}\left(f_{\Theta_i^\prime}, \mathcal{Q}_i\right)$
        
    \ENDFOR
\end{algorithmic}
\end{algorithm}

\section{Experimental Setups}

\subsection{Datasets}
\label{sec: datasets}
We use the python package \textit{pyroomacoustics} \cite{pyroomacoustics} to simulate shoebox-shaped rooms and utilize the computationally simulated SRIRs to synthesize the datasets containing 30,000 5-second clips with reverberation time (RT60) from 0.3 seconds to 0.5 seconds. Sound event examples from FSD50K and AudioSet are cleaned by the pre-trained CNN14\cite{kong2020panns} model. We utilize CNN14 to infer sound event samples and select high-quality samples based on the output probability. The maximum polyphony of target classes is 3, excluding the additional polyphony of interference classes. We refer to the computationally simulated datasets as CSD. For simple comparison and reproducibility, we adopt the official synthetic datasets \cite{official}, which are synthesized using collected SRIRs for the baseline training of Task 3 of the DCASE Challenge in 2022 and 2023 \cite{starss22,starss23}. The official synthetic datasets are denoted as \textit{Base Dataset} or \textit{Base} in this work. The Base Dataset contains 1200 1-minute audio clips, with a maximum polyphony of 2 and no directional interference, and is synthesized using real-scene SRIRs from TAU-SRIR DB \cite{srir-db}, which are measured in 9 rooms at Tampere University. The STARSS23 \cite{starss23} dataset, an extended version of the STARSS22 \cite{starss22} dataset, is also recorded in real-world environments and annotated manually. STARSS23 encompasses all recordings of STARSS22 and includes additional 4-hour recordings. There are 16 different recording rooms in total in the development set of the STARSS23 dataset, including nine recording rooms in \textit{dev-train-set} and seven recording rooms in \textit{dev-test-set}. The labels of the STARSS23 evaluation set remain inaccessible, and hence all subsequent ablation experiments are based solely on the analysis of the STARSS23 development set. To further validate the effectiveness of our proposed method, we present the performance of the computationally synthesized scenes in Section \ref{sec: exp. sim_scene}. We denote the computationally synthesized scenes as CSS. The composition of all synthetic datasets we used in this work is illustrated in Table \ref{tab: dataset_sec}.

\begin{table}
    \centering
    \caption{Composition of synthetic datasets.}
    \begin{adjustbox}{width=\columnwidth,center}
    \begin{tabular}{ccc}
        \toprule[1pt]
        Name & Sound events & SRIRs \\
        \midrule
        Base & FSD50K* & \makecell[l]{TAU-SRIR DB} \\
        CSD & FSD50K\&AudioSet & \makecell[l]{Computationally Simulated SRIRs \\ based on shoebox-shaped room simulation}\\
        CSS & FSD50K\&AudioSet & \makecell[l]{Computationally Simulated SRIRs \\ based on 3D mesh rooms simulation}\\
        \bottomrule[1pt]
    \end{tabular}
    \end{adjustbox}
    \label{tab: dataset_sec}
    \begin{justify}
        *FSD50K includes no sound event labelled as \textit{Background and pop music}, which is sourced from the public domain for supplementation \cite{official}.
    \end{justify}
\end{table}

\begin{figure}[t!]
    \centering
    \includegraphics[width=\linewidth]{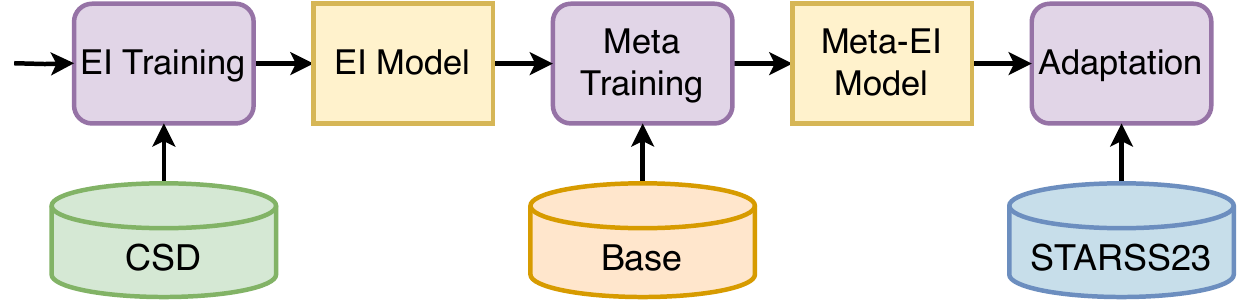}
    \caption{A flowchart showing training procedures of our proposed method, from a randomly initialized model to the final model to adapt to unknown environments. The cylinders denote the dataset we used in different stages of the training.}
    \label{fig: meta-seld++ flowchart}
\end{figure}

The environment-adaptive Meta-SELD involves a three-stage pipeline, each stage utilizing a distinct dataset, as illustrated in Fig. \ref{fig: meta-seld++ flowchart}. We start by randomly initializing the EI model and training it on CSD. Subsequently, selective memory meta-learning is applied to the EI model on the Base Dataset, served as the meta-training set $\mathcal{D}_\text{train}$, to create the meta-EI model. Finally, fast adaptation is performed to a specific environment from STARSS23 using a few samples recorded in the corresponding environment. All development sets of STARSS23 are used for meta-test set $\mathcal{D}_\text{test}$ to evaluate the performance of the adaptation to unknown environments.

$\mathcal{D}_\text{train}$ and $\mathcal{D}_\text{test}$ are divided into 9 tasks and 16 tasks, respectively, corresponding to 9 rooms and 16 rooms. In the meta-training phase, we first sample a batch of rooms and then sample a batch of examples from each room. A batch of samples from an individual room constructs a task, and a part of the samples are support samples while the remaining samples are query samples. The batch sizes of rooms and samples are set to 9 and 128, respectively. The batch of rooms consists of all 9 rooms sampled from $\mathcal{D}_\text{train}$ without replacement. For each batch of samples, the first 30 samples are designated as support samples in this work unless specified otherwise. In the meta-testing phase, samples from each room, excluding the support samples from $\mathcal{S}_j^\mathtt{test}$, comprise the final test (query) set $\mathcal{Q}_j^\mathtt{test}$. We will evaluate the performance of parameterized function $f_{\Theta_j^\prime}$ with adaptation parameters $\Theta_j^\prime$ on $\mathcal{Q}_j^\mathtt{test}$ after iteration updates on $\mathcal{S}_j^\mathtt{test}$ with the initial parameters $\Theta$.

\subsection{Evaluation metrics}
\label{sec: metric}
We use a joint metric of localization and detection\cite{mesaros2019joint,overviewofDCASE} here: two location-dependent detection metrics, F-score ($\mathrm{F}_{\leq 20^\circ}$) and error rate ($\mathrm{ER}_{\leq 20^\circ}$), and two class-dependent localization metrics, localization recall ($\mathrm{LR}_\mathrm{CD}$) and localization error ($\mathrm{LE}_\mathrm{CD}$). $\mathrm{F}_{\leq 20^\circ}$ and $\mathrm{ER}_{\leq 20^\circ}$ consider true positives predicted within a spatial threshold $20^\circ$ away from the ground truth. $\mathrm{LE}_\mathrm{CD}$ and $\mathrm{LR}_\mathrm{CD}$ compute the mean angular error and true positive rate in the case when the types of sound events are predicted correctly, respectively.

We use an aggregated SELD metric for the method comparison and hyper-parameter selection: 
\begin{equation}
\scalebox{0.9}{$
{\mathcal{E}_\mathtt{SELD}}=\frac{1}{4}\left[\mathrm{ER}_{\leq 20^{\circ}}+\left(1-\mathrm{F}_{\leq 20^{\circ}}\right)+\frac{\mathrm{LE}_{\mathrm{CD}}}{180^{\circ}}+\left(1-\mathrm{LR}_{\mathrm{CD}}\right)\right].$}
\end{equation}

A macro-average of $\mathrm{F}_{\leq 20^\circ}$, $\mathrm{LR}_\mathrm{CD}$, $\mathrm{LE}_\mathrm{CD}$, and $\mathcal{E}_\mathtt{SELD}$ across classes is used. A good system should have small $\mathrm{ER}_{\leq 20^{\circ}}$, large $\mathrm{F}_{\leq 20^{\circ}}$, small $\mathrm{LE}_{\mathrm{CD}}$, large $\mathrm{LR}_{\mathrm{CD}}$, and small ${\mathcal{E}_\mathtt{SELD}}$. Note that different from \cite{meta-seld}, where room-wise metrics are micro-averaged in the end, we compute the metrics in each room and then macro-average the metrics across rooms.

\subsection{Hyper-parameters}
The sampling rate is 24 kHz. We extract 64-dimensional log mel spectrograms from four-channel FOA signals with a Hanning window of 1024 points and a hop size of 320. Each audio clip is segmented to a fixed length of five seconds with no overlap for training and inference. 

AdamW \cite{adamw} is used to update meta parameters, while SGD is used to update adaptation parameters. The batch size is 128. For training the conventional supervised learning models, the learning rate is set to 0.001 for the first 60 epochs out of 80 epochs and is then decreased by 10\% every 10 epochs. In the meta-training phase, we find that setting the momentum to 0.01 in the Batch Normalization layer \cite{batchnorm} results in more stable training and better performance. We set one epoch containing nine meta batches, which encompass all rooms within the Base Dataset. Subsequently, the gradients of one epoch are averaged to update meta parameters in the outer loop step. For training Meta-SELD++, the learning rate is 0.0003 for the first 300 epochs out of 500 epochs and is then decreased by 10\% every 100 epochs. If not specified, only 5 update steps are considered in the inner loop of MAML in subsequent experiments. All networks are implemented using PyTorch.
 
\section{Experiments}
\subsection{Effect of synthetic data}
\label{sec: exp. synth_data}

\begin{table}[t!]
    \centering
    \caption{Results of the computationally synthesized datasets. Methods are evaluated on all development sets of STARSS23.}
        \begin{tabular}{c|cccc|c}
            \toprule[1pt]
            Dataset& $\mathrm{ER}_{20^{\circ}}\downarrow$ & $\mathrm{F}_{20^{\circ}}\uparrow$ &
            $\mathrm{LE}_\mathrm{CD}\downarrow$ & $\mathrm{LR}_\mathrm{CD}\uparrow$ & $\mathcal{E}_{\mathtt{SELD}}\downarrow$\\
            \midrule 
             Base& 0.722& \textbf{23.2\%}& $\mathbf{22.2^\circ}$& 39.5\%& 0.555  \\
             CSD & 0.746& 20.9\%& $25.6^\circ$& 41.4\%& 0.566 \\
             Base + CSD& \textbf{0.697}& 23.1\%& $22.6^\circ$& \textbf{44.3\%}&  \textbf{0.537}\\
             \bottomrule[1pt]
        \end{tabular}
    \label{tab: CSD}
\end{table}

We evaluate our synthetic data using the conventional supervised learning method. The model is trained on the Base Dataset (Base) and CSD, and then evaluated on all development sets of STARSS23. 

Table \ref{tab: CSD} shows the results of the data synthesis method. The results demonstrate that the model trained on the CSD can generalize to real-scene datasets. Comparing Base with CSD, we observe the performance gap is mainly in localization. One of the possible reasons is that there is some discrepancy between computationally simulated SRIRs and measured SRIRs in real scenes. In addition, adding CSD to the Base Dataset for training further improves the performance, particularly in detection, perhaps because CSD increases the diversity of sound events.

\subsection{Effect of Meta-SELD}

\begin{table}
    \centering
    \caption{Description of utilized models. ($\cdot$) denotes the dataset for training. $\mathcal{S}_j^\text{test}$ indicates the support set from the $j$-th room of STARSS23.}
    \begin{adjustbox}{width=\columnwidth,center}
    \begin{tabular}{llc}
        \toprule[1pt]
        Approach & Initial parameters& Datasets \\
        \midrule
        \textbf{\makecell[l]{Conventional supervised \\ learning methods}}\\
        SELD ($\cdot$) &\makecell[c]{-} & \makecell[l]{The combinations of \\ Base, CSD, and $\mathcal{S}^\mathtt{test}_j$}  \\
        \makecell[l]{SELD ($\cdot$) w/ adapt. or\\ Fine-tuned SELD} & \makecell[l]{SELD (Base) or\\ SELD (Base + CSD)} & $\mathcal{S}^\mathtt{test}_j$ \\ 
        \midrule
        \textbf{Meta learning methods}\\
        Meta-SELD &\makecell[c]{-} & Base \\
        Meta-SELD w/ adapt. &Meta-SELD & $\mathcal{S}^\mathtt{test}_j$ \\
        Meta-SELD++ &SELD (CSD) & Base \\
        Meta-SELD++ w/ adapt. &Meta-SELD++ & $\mathcal{S}^\mathtt{test}_j$ \\
        Sel. Mem. Meta-SELD* &SELD (CSD) & Base \\
        Sel. Mem. Meta-SELD w/ adapt. &Sel. Mem. Meta-SELD & $\mathcal{S}^\mathtt{test}_j$ \\
        \bottomrule[1pt]
    \end{tabular}
    \end{adjustbox}
    \begin{justify}
        *Selective Memory Meta-SELD with environment representations is referred to as environment-adaptive Meta-SELD.
    \end{justify}
    \label{tab: model_desc}
\end{table}

To demonstrate the effectiveness of our proposed Meta-SELD, we compare the Meta-SELD, the SELD method, and the fine-tuned SELD method. The differences among these methods are described in Table \ref{tab: model_desc}. Macro-averaged metrics for all 16 rooms of STARSS23 are shown in Table \ref{tab: SELDvs.Meta-SELD}. In SELD (Base + $\mathcal{S}_j^\text{test}$), the support set $\mathcal{S}_j^\text{test}$ from the $j$-th room of STARSS23 is added to the synthetic datasets for training from scratch, and the query set $\mathcal{Q}_j^\text{test}$ is used for evaluating that specific model. This approach, however, requires training multiple models from scratch, one for each specific room of STARSS23. In SELD (Base), we first train an EI model on the Base Dataset and then fine-tune (adapt) on $\mathcal{S}_j^\text{test}$. In Meta-SELD, we apply MAML to a SELD model with random initialized meta-parameters. The SELD (Base) and Meta-SELD methods without adaptation refer to no fine-tuning on $\mathcal{S}_j^\text{test}$. 

The top block of Table \ref{tab: SELDvs.Meta-SELD} presents the method using the Base Dataset for training or meta-training, while the bottom block presents the method using the Base Dataset and computationally simulated datasets. In terms of the average performance, we can see that after adaptation on $\mathcal{S}_j^\text{test}$, both Meta-SELD and Meta-SELD++ exhibit superior performance in comparison to the corresponding fine-tuned SELD method. In a comparison between SELD (Base) and Meta-SELD, we observe a bigger performance gap in Meta-SELD with and without adaptation, which means the meta parameters of Meta-SELD are more suitable for adapting to a new environment. However, the performance gap in Meta-SELD++ becomes smaller and the meta parameters from Meta-SELD++ outperform the parameters from SELD (Base + CSD). We conjecture the phenomenon results from conflicts while optimizing among environments, which is especially obvious in Meta-SELD++. The SELD method, which adds the support set $\mathcal{S}_j^\text{test}$ for training, performs the best among these methods. This phenomenon may be attributed to the fact that training multiple independent models from scratch could avoid compromise in these environments.

Meanwhile, we consider that, compared to Meta-SELD++, the experimental results of Meta-SELD are more appropriate to represent the ability for fast adaptation, which refers to relative performance improvement, the performance difference between with and without adaptation. However, Meta-SELD++ can improve absolute performance or performance without adaptation while concurrently preserving the ability for fast adaptation based on the pre-trained EI model.

\begin{table}[t!]
    \centering
    \caption{The performance of meta-learning-based and conventional supervised-learning-based methods. All of the methods are evaluated on $\mathcal{Q}_i^\text{test}$ from STARSS23.}
    \label{tab: SELDvs.Meta-SELD}
    \begin{adjustbox}{width=\columnwidth,center}
        \begin{tabular}{c|c|cccc|c}
            \toprule[1pt]
             Method&  Adapt. &$\mathrm{ER}_{20^{\circ}}\downarrow$ & $\mathrm{F}_{20^{\circ}}\uparrow$ &
             $\mathrm{LE}_\mathrm{CD}\downarrow$ & $\mathrm{LR}_\mathrm{CD}\uparrow$ & $\mathcal{E}_{\mathtt{SELD}}\downarrow$\\
             \midrule
             SELD (Base + $\mathcal{S}_i^\text{test}$) & -& 0.642& 32.8\%& $21.1^\circ$& 54.5\% & 0.471\\
             \midrule
             \multirow{2}{*}{SELD (Base)}& \XSolidBrush & 0.742& 30.1\%& $23.9^\circ$& 53.6\%& 0.509 \\
             &  \CheckmarkBold & 0.690& 31.6\%& $22.8^\circ$& 53.7\%& 0.491 \\
             \midrule
             \multirow{2}{*}{Meta-SELD}& \XSolidBrush & 0.718& 29.8\% & $24.0^\circ$& 53.5\%& 0.504 \\
             &  \CheckmarkBold & 0.639& 32.6\% & $21.3^\circ$& 52.8\%& 0.476 \\
             \midrule
             \midrule
             \makecell{SELD \\(Base + CSD + $\mathcal{S}_i^\text{test}$)}& -& 0.653& 33.6\%& $22.6^\circ$& 58.2\%& 0.465\\
             \midrule
             \multirow{2}{*}{SELD (Base + CSD)}& \XSolidBrush & 0.714& 28.6\%& $25.2^\circ$& 58.8\%& 0.495 \\
             &  \CheckmarkBold & 0.686& 30.1\%& $24.7^\circ$& 59.2\%& 0.482 \\
             \midrule
             \multirow{2}{*}{Meta-SELD++}& \XSolidBrush & 0.698& 30.5\%& $23.6^\circ$& 59.6\%& 0.482 \\
             &  \CheckmarkBold & 0.657& 31.4\%& $23.2^\circ$& 59.0\%& 0.470 \\
            \bottomrule[1pt]
        \end{tabular}
    \end{adjustbox}
\end{table}

\subsection{Effect of selective memory}
\label{sec: exp. sel_mem}
\subsubsection{Inputs to selective memory}

\begin{figure}[t]
  \centering
  \centerline{\includegraphics[width=\columnwidth]{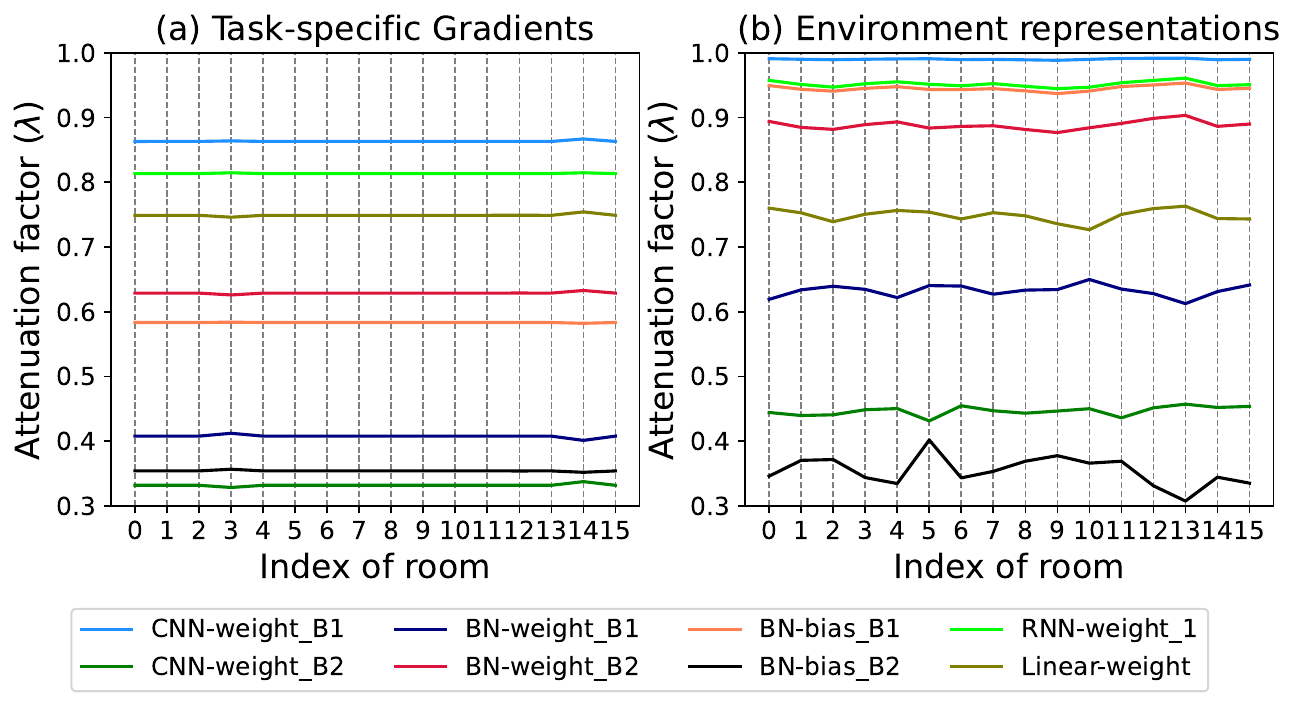}}
  \caption{Generated room-and-layer-wise attenuation factors $\lambda$ using selective memory for each room of STARSS23.}
  \label{fig: l2f-lambda}
\end{figure}

\begin{table*}[t]
    \centering
    \caption{Ablation studies on types of input in selective memory Meta-SELD. The $\pm$ shows 95\% confidence intervals.}
    \begin{adjustbox}{center}
    \begin{tabular}{c|c|cccc|c}
    \toprule[1pt]
    Method & Input &$\mathrm{ER}_{20^{\circ}}\downarrow$ & $\mathrm{F}_{20^{\circ}}\uparrow$ &             $\mathrm{LE}_\mathrm{CD}\downarrow$ & $\mathrm{LR}_\mathrm{CD}\uparrow$ & $\mathcal{E}_{\mathtt{SELD}}\downarrow$\\
    \midrule
    \makecell{Meta-SELD++ w/ adapt.} & \multirow{1}{*}{-}& \multirow{1}{*}{0.657}& \multirow{1}{*}{31.4\%}& \multirow{1}{*}{$23.2^\circ$}& \multirow{1}{*}{59.0\%}& \multirow{1}{*}{0.470} \\
    \midrule
    \multirow{3}{*}{\makecell{+ Selective Memory}} & None& 0.633 $\pm$ 0.002 & 33.9\% $\pm$ 0.4\% & $21.4^\circ \pm 0.1^\circ$& 60.0\% $\pm$ 0.7\%& 0.453 $\pm$ 0.002\\
    & Gradients & 0.636 $\pm$ 0.005& 34.7\% $\pm$ 0.4\% & $21.3^\circ \pm 0.3^\circ$& 57.7\% $\pm$ 0.4\%& 0.458 $\pm$ 0.003\\
    & {Representations}& 0.632 $\pm$ 0.007& 34.9\% $\pm$ 1.1\% & $21.6^\circ \pm 0.3^\circ$& 60.0\% $\pm$ 0.8\%& 0.451 $\pm$ 0.006\\
    \bottomrule[1pt]
    \end{tabular}
    \end{adjustbox}
    \label{tab: Meta-Pretrain+L2F}
\end{table*}

Table \ref{tab: Meta-Pretrain+L2F} shows the results of three types of input in the selective memory methods, the task-specific gradients on the support set $\mathcal{S}_i$, "None", and the environment representations. The input "None" in the selective memory method refers to layer-wise learnable parameters as attenuation coefficients instead of being generated by the attenuation network. We see that applying selective memory to Meta-SELD++ obtains performance improvement, particularly in localization. These methods are even more effective than the SELD (Base + CSD + $\mathcal{S}_i^\text{test}$) in Table \ref{tab: SELDvs.Meta-SELD}. Among these types of inputs, the environment representations perform better. This exhibits the effectiveness of environment representations as the input to the selective memory method.

\subsubsection{Attenuation factors}

Fig. \ref{fig: l2f-lambda} illustrates the attenuation factor $\lambda$ of a few typical layers for each room of STARSS23. However, observing room-and-layer-wise attenuation factor $\lambda$ derived from the task-specific gradients, we note that $\lambda$ varies over a small range from room to room. This suggests that $\lambda$ can not be adaptive to the diverse acoustic environments of these rooms. We analyze the inputs to the attenuation network, task-specific gradients on $\mathcal{S}_i$, and find that most gradients exhibit diminutive magnitudes. This phenomenon can be attributed to the pre-trained EI model, which initializes the meta-parameters of Meta-SELD++, resulting in minute gradient values. In contrast, a comparison of the gradients and representations as input to the attenuation network indicates that environment representations generate more environment-adaptive attenuation factors $\lambda$. The changes in these attenuation factors are more conspicuous from room to room.

We adopt a typical CNN block architecture, Conv2d-BatchNorm2d-ReLU. Fig. \ref{fig: l2f-lambda} presents the attenuation factors of the shallow-layer CNN block and deep-layer CNN block, denoted as B1 and B2, respectively. Given that the attenuation coefficients of Conv2d and BatchNorm2d act on a CNN block together, it is observed that generated $\lambda$ from B2 is more sensitive to environmental variations than that from B1. This suggests that the deep layers prefer environment-adaptive attenuation and encode environment-dependent features, which aligns with the observation of \cite{l2f}.

\begin{figure*}[t]
    \centering
    \includegraphics[width=1\textwidth]{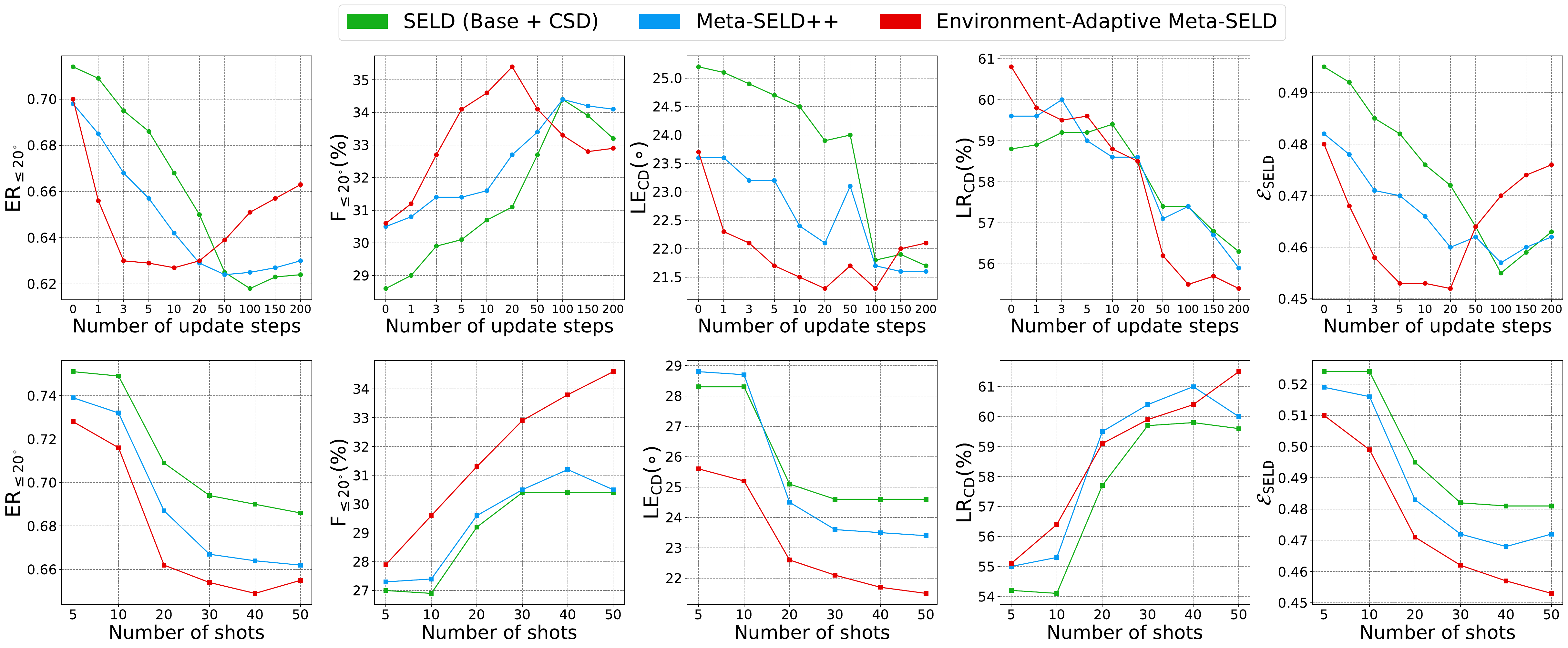}
    \caption{The effect of hyper-parameters of MAML. Each column represents one metric. \textbf{Top row}: The number of inner-loop update steps. \textbf{Bottom row}: The number of support samples.}
    \label{fig:steps_shots}
\end{figure*}

\subsubsection{Representations of Environments}
\begin{table*}[t]
    \centering
    \caption{Ablation studies on approaches for extracting environment representations.}
    \begin{adjustbox}{center}
    \begin{tabular}{c|cccc|c}
    \toprule[1pt]
    Approach & $\mathrm{ER}_{20^{\circ}}\downarrow$ & $\mathrm{F}_{20^{\circ}}\uparrow$ &             $\mathrm{LE}_\mathrm{CD}\downarrow$ & $\mathrm{LR}_\mathrm{CD}\uparrow$ & $\mathcal{E}_{\mathtt{SELD}}\downarrow$\\
    \midrule
    Last embeddings (Mean)& 0.668 $\pm$ 0.040& 32.1\% $\pm$ 3.8\%& $23.0^\circ \pm 1.6^\circ$& 57.2\% $\pm$ 3.7\%& 0.476 $\pm$ 0.030 \\
    Last embeddings (Encode) & 0.635 $\pm$ 0.009& 35.3\% $\pm$ 2.1\%& $21.4^\circ \pm 0.4^\circ$& 59.3\% $\pm$ 1.3\%& 0.452 $\pm$ 0.009\\
    Environment representations & 0.632 $\pm$ 0.007& 34.9\% $\pm$ 1.1\% & $21.6^\circ \pm 0.3^\circ$& 60.0\% $\pm$ 0.8\%& 0.451 $\pm$ 0.006\\
    \bottomrule[1pt]
    \end{tabular}
    \end{adjustbox}
    \label{tab: L2F-emb}
\end{table*}

We investigate various techniques for extracting environment representations: feature maps from the last layer (before the linear layer) that are averaged in the batch and time axes, feature embeddings derived from the last layer, and environment representations constituted by concatenated feature embeddings from all preceding layers. Table \ref{tab: L2F-emb} demonstrates the effectiveness of the environment extractor and preceding feature embeddings. Feature maps from the last layer encoded by the environment extractor perform better compared to being directly averaged. Moreover, preceding feature embeddings provide more information and further performance improvement. 

\begin{figure*}[t]
  \centering
  \scalebox{1.0}{\centerline{\includegraphics[width=\textwidth]{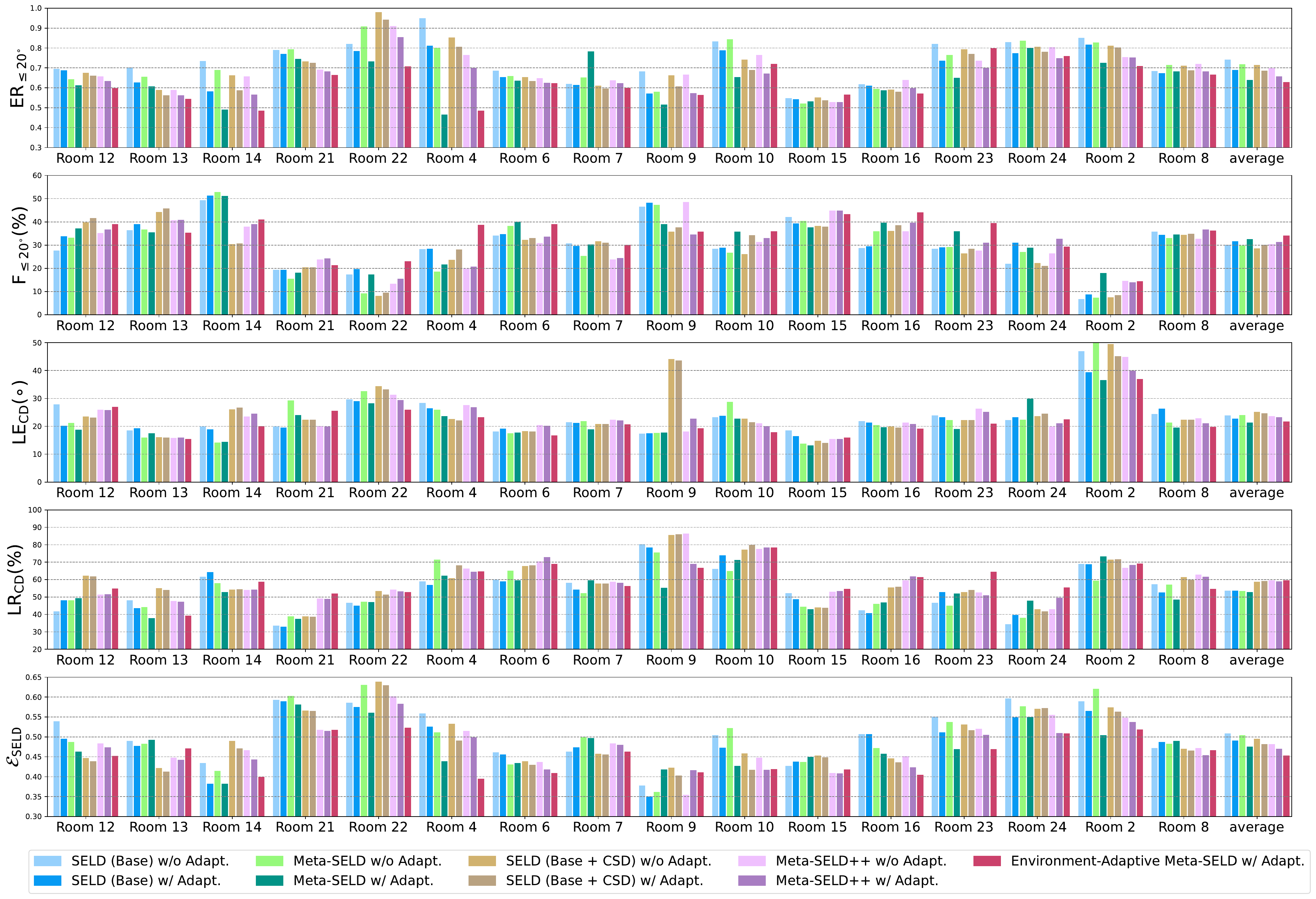}}}
  \caption{Room-wise performance of all development sets of STARSS23. All methods are evaluated on $\mathcal{Q}_j^\text{test}$. Note that the \textit{average} score of each metric computes the macro-average performance of all these rooms.}
  \label{fig: score_aggregation}
\end{figure*}

\subsection{Effect of adaptation setups}

The most important hyper-parameters in the adaptation phase of MAML include the number of inner-loop update steps and support samples. Intuitively, the inner-loop optimization should be consistent during meta-training and meta-testing. However, the large number of update steps and support samples leads to excessive computation and memory burdens. We exploit previously well-trained parameters and investigate the effect of adaptation setups in meta-testing. 

The top row of Fig. \ref{fig:steps_shots} shows the effect of the number of update steps on the adaptation of SELD (Base + CSD), Meta-SELD++, and environment-adaptive Meta-SELD. The number of support samples is 30. Experimental results demonstrate that meta-learning-based methods exhibit faster adaptation than the conventional supervised-learning-based method. The environment-adaptive Meta-SELD achieves more efficient adaptation. The benefits come from selectively memorizing necessary information, helping the learner adapt to new environments more quickly. Also, we observe Meta-SELD++ and SELD (Base+CSD) achieve similar performance after exceeding 100 update steps. This is likely because the initialized parameters of Meta-SELD++ are derived from the pre-trained SELD (CSD), resulting in minimal differences between their respective parameters. After extensive fine-tuning over numerous update steps, both sets of fine-tuned (adaptation) parameters undergo over-fitting and exhibit similar convergence.

We also investigate the effect of the number of shots (support samples). We select the first 50 samples of each room as the support set, and all remaining samples are as the query set. We set the number of update steps to 5 for consistency. The bottom row of Fig. \ref{fig:steps_shots} shows that meta-learning-based methods exploit support samples more effectively. When the number of support samples increases in the environment-adaptive Meta-SELD method, the performance is consistently improved, but the magnitude of performance improvement also appears to decrease.

\subsection{Room-wise performance}

Fig. \ref{fig: score_aggregation} shows the room-wise metrics. Through a comparative analysis of conventional supervised-learning-based and meta-learning-based methods on identical datasets, we observe that meta-learning-based methods can reduce $\mathcal{E}_\mathtt{SELD}$ effectively in rooms where conventional supervised-learning-based methods exhibit high $\mathcal{E}_\mathtt{SELD}$, such as Room 2, Room 4, and Room 22. This means when a model has an unsatisfactory generalization to a specific room below the average performance, using a few samples collected in the specific room for adapting could improve the performance significantly. However, we also observe performance degradation or insignificant improvement in $\mathcal{E}_\mathtt{SELD}$, even though some new samples of unseen environments are used for adapting in both conventional supervised-learning-based and meta-learning-based methods, for example, in Room 7 and Room 15. This phenomenon could arise from the fact that our methods have difficulty extracting valid information for training from new samples. In Room 9, the performance of SELD (Base) and SELD (Base + CSD) is improved after fine-tuning using some new samples of the corresponding room, but the performance of Meta-SELD and Meta-SELD++ after adaptation is degraded. The reason may be that compromised initial meta parameters among those rooms fail to adapt to the environment. The meta-learning-based methods find general initial parameters that can be adapted to unknown environments in the sense of average, and experimental results demonstrate that meta-learning-based methods outperform the conventional supervised-learning-based methods in most rooms and perform better on average. 

In contrast to SELD (Base + CSD), the environment-adaptive Meta-SELD exhibits comparable average performance in $\mathrm{LR}_\mathtt{CD}$, but lower $\mathrm{LE}_\mathtt{CD}$. Consequently, performance improvement in localization is the main factor for the reduction in $\mathcal{E}_\mathtt{SELD}$. Additionally, the performance improvement in Room 2 and Room 22 can also be attributed to analogous factors.

\subsection{Environment representations}
\label{sec: exp. env_rep}

\begin{figure*}[t!]
    \begin{center}
        \subfloat{
        \includegraphics[width=0.235\linewidth]{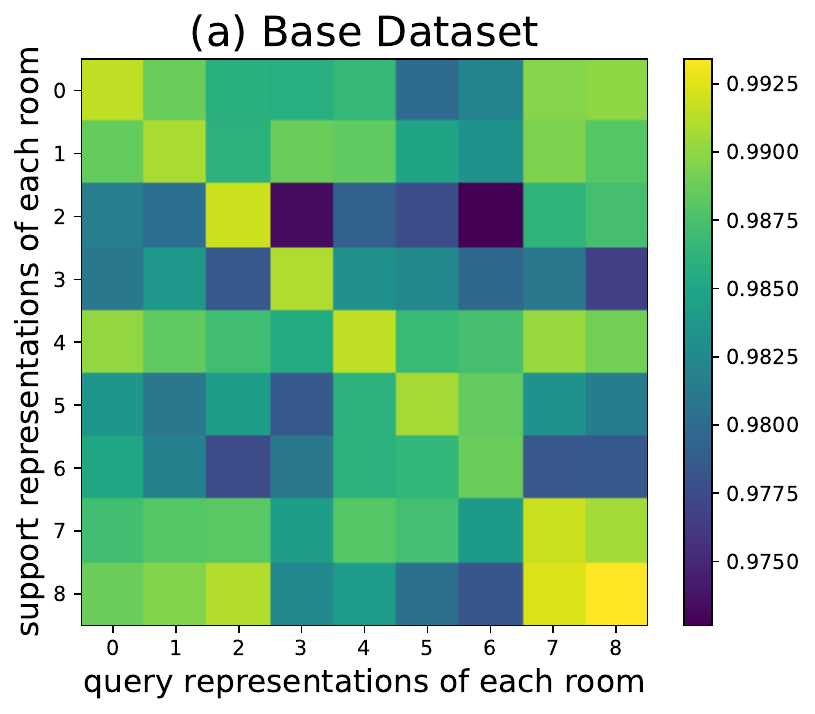}
        \label{fig: sim_official}
        }
        \subfloat{
        \includegraphics[width=0.235\linewidth]{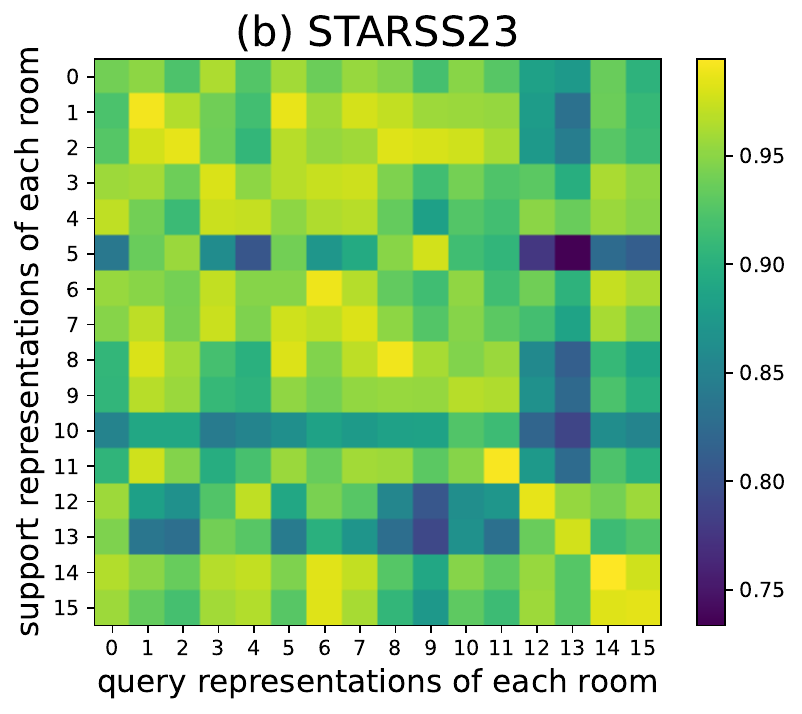}
        \label{fig: sim_starss23}
        }
        \subfloat{
        \includegraphics[width=0.235\linewidth]{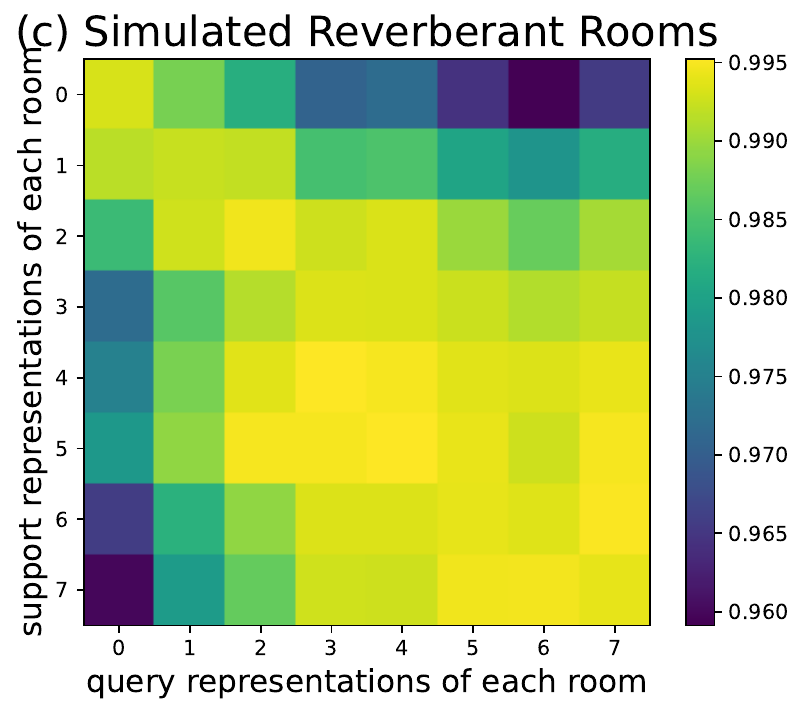}
        \label{fig: sim_reverb}
        }
        \subfloat{
        \includegraphics[width=0.235\linewidth]{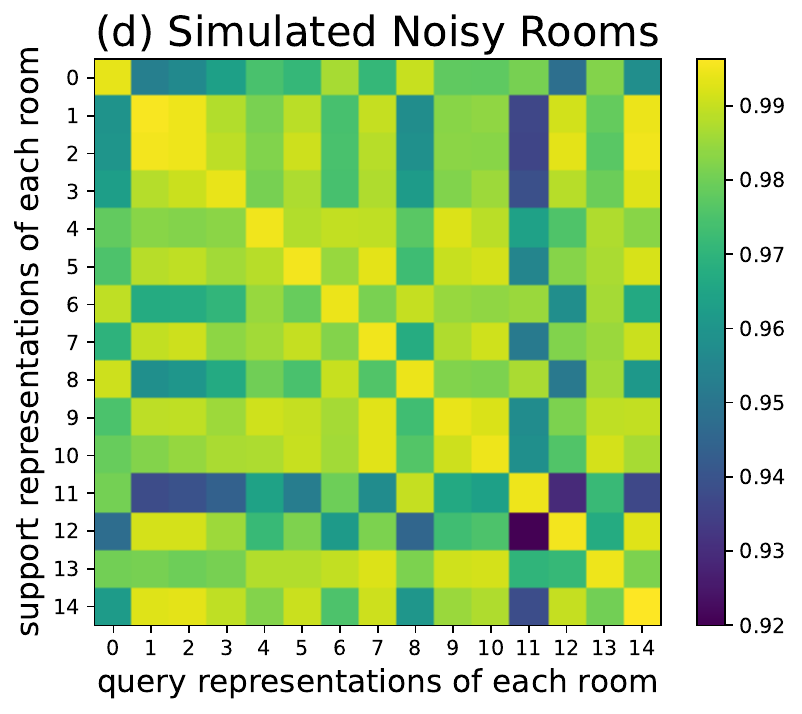}
        \label{fig: sim_noise}
        }
    \end{center}

    \begin{center}
        \subfloat{
        \includegraphics[width=0.235\linewidth]{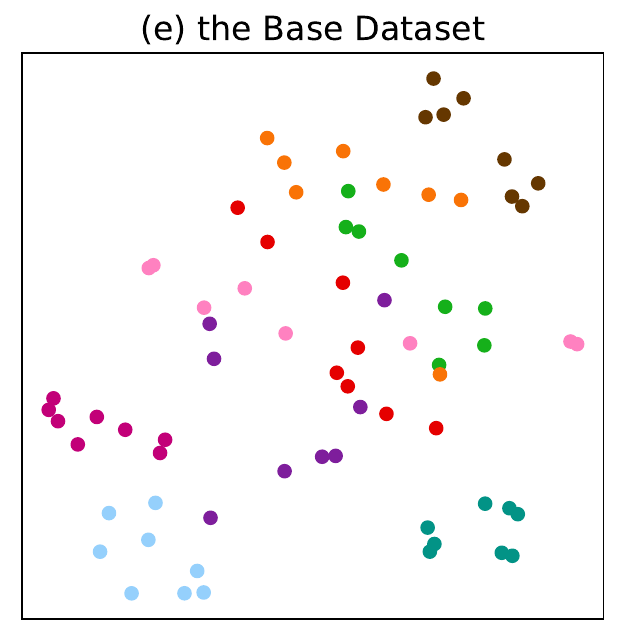}
        \label{fig: tsne_official}
        }
        \subfloat{
        \includegraphics[width=0.235\linewidth]{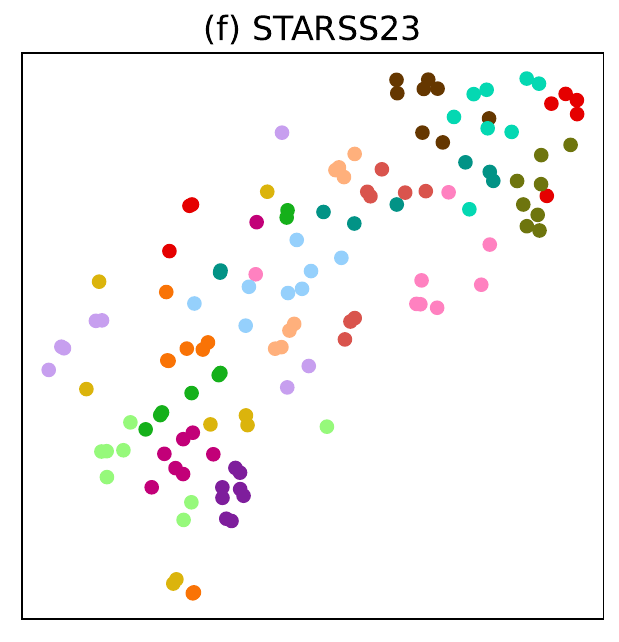}
        \label{fig: tsne_starss23}
        }
        \subfloat{
        \includegraphics[width=0.235\linewidth]{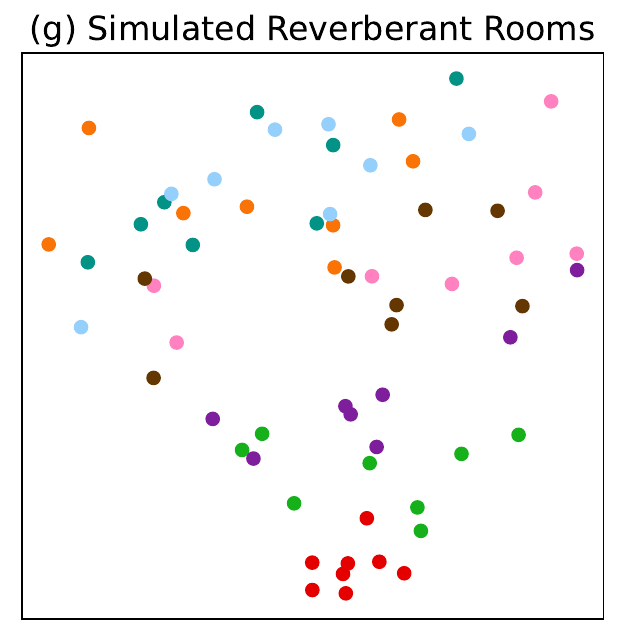}
        \label{fig: tsne_reverb}
        }
        \subfloat{
        \includegraphics[width=0.235\linewidth]{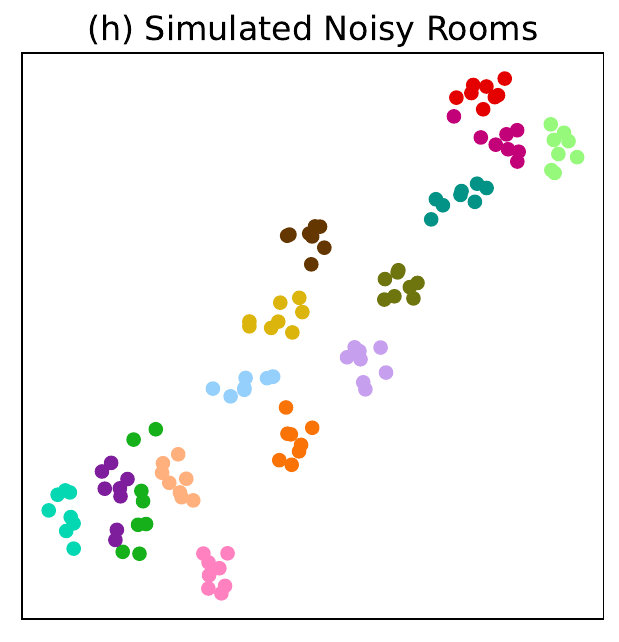}
        \label{fig: tsne_noise}
        }
    \end{center}
    \caption{Visualization of environment representations. The four plots (a-d) of the top row display the cosine similarity maps between support representations and query representations. The four plots (e-h) of the bottom row display the representations via t-SNE, and points of the same color indicate representations from the same room. Four datasets are shown: the Base Dataset, the development set of STARSS23, computationally simulated reverberant rooms, and computationally simulated noisy rooms.}
    \label{fig: visualization}
\end{figure*}

To interpret the representation extracted by the sub-network in Fig. \ref{fig: maml+l2f-crnn}, we study the relationship between the representation and the environment. We hypothesize that different clips recorded at various spatial positions in the same environment generally have more similar acoustic properties in contrast to various environments, except in extreme cases. Therefore, we extract representations from several disjoint batches of samples within the same dataset, and then we compute the similarity between the representations and visualize these representations via t-SNE \cite{t-sne}. Fig. \ref{fig: visualization} shows the visualization of extracted representations.

The SELD performance is greatly affected by noise and reverberation. In our study, we simulate environments with varying acoustic properties, including reverberant rooms and noisy rooms, to investigate how the representations are affected by different acoustic properties. For the reverberant simulations, rooms of the same size are simulated at intervals of 0.3-second RT60, ranging from 0.4 to 2.5 seconds, therefore, there are 8 rooms with different RT60 in total. For the noisy simulations, we simulate 15 rooms of the same size, absorption coefficients, and reflection orders. Subsequently, we add ambient noise from the NoiseX-92 \cite{noisex92} database into synthetic recordings to generate SNR ranging from 10 dB to 15 dB. The NoiseX-92 database contains 15 types of ambient noise. Therefore, each room has a unique noise type. All sound event examples for synthesizing are randomly sampled from FSD50K and AudioSet. Based on the previously trained model using the environment-adaptive Meta-SELD method, we directly perform an inference (meta-test) on these simulated datasets.

\subsubsection{Similarity Maps}
Representations from the same or similar rooms should have a high similarity. In this work, the cosine distance of two representations is used to measure their similarity. For a batch of 128 samples from the same room, we extract support representations from the first 30 samples, corresponding to the number of support samples for adaptation, and query representations from the last 98 samples. The cosine similarity is computed between support representations from a room and query representations from the same or other rooms. 

Fig. \ref{fig: visualization} (a) and (b) present the cosine similarity maps on the Base Dataset and STARSS23. The experimental results show that all diagonal elements are the maximum of their rows of the similarity matrix on the Base Dataset, while $12/16$ of diagonal elements are the maximum of their rows on STARSS23. Fig. \ref{fig: visualization}(c) and (d) show the cosine similarity maps on the simulated reverberant rooms and simulated noisy rooms. We see that these maps exhibit a notable degree of symmetry. The diagonal elements of the similarity matrix are pretty large. 

Notably, the cosine distance values in the similarity map appear to be fairly large, typically above 0.9. This phenomenon may be attributed to the extracted representations containing information about the characteristics of microphones and background noise in STARSS23 and the Base dataset or to significant distribution differences between the Base dataset and the simulated datasets in the simulated reverberant and noisy rooms. However, the environment extractor may still implicitly learn representations relevant to environments and recognize different environments. In addition, we observe environment representations have a high resolution in rooms with low RT60, but a low resolution in rooms with high RT60. The differences in the environment representations of these rooms with high RT60 are small. This may result from the small range RT60 of the training set. 

\subsubsection{Visualization of representations via t-SNE}

We sample 8 batches of clips from each dataset, with 32 clips per batch, and then compute representations for each batch\footnote{For STARSS23, the number of clips is less than 256 in a few rooms, therefore, we repeat the corresponding batch of clips along the batch dimension.}. Empirically, we find the learned representations cluster meaningful: representations from the same room tend to be clustered, as shown in Fig. \ref{fig: visualization} (e-h). Especially in Fig. \ref{fig: visualization}(h), different recordings with the same noisy type have similar features, which leads to better clustering performance.

These observations demonstrate the extracted representations are relevant to the environments.  

\subsection{Results on computationally synthesized scenes}
\label{sec: exp. sim_scene}

We evaluate the method on computationally synthesized scenes to further validate the effectiveness of environment-adaptive Meta-SELD. These scenes are represented as semantically labeled 3D meshes from the 3D-FRONT dataset \cite{3d-front}, which contains 18,968 diversely furnished rooms in 6,813 professionally designed houses. We computationally simulate SRIRs of 15 houses sampled from the 3D-FRONT dataset, which is the same as the simulation of the Geometric-Wave Acoustic (GWA) dataset \cite{GWA}. We synthesize 256 5-second spatial recordings for each acoustic environment with these simulated SRIRs and sound events from FSD50K \cite{fonseca2021fsd50k} and AudioSet \cite{gemmeke2017audio}. Ambient noise from the NoiseX-92 \cite{noisex92} database is also mixed into spatial recordings to generate signal-to-noise (SNR) ranging from 10 dB to 15 dB.

We leverage the previous methods trained on CSD and Base Dataset to perform an inference (meta-test) on the simulated dataset. Table \ref{tab: simulated scene} presents the results on computationally synthesized scenes. Notably, we see weak generalizations to these environments for both conventional supervised learning and meta-learning methods, due to disparities in environmental conditions between the training and test datasets. Additionally, the considerable variation in distribution between meta-training sets and meta-test sets puts the meta-learned prior knowledge at high risk of losing effectiveness \cite{closerlook, id_ood, secure_ood}. However, the incorporation of selective memory into Meta-SELD++ manifests superior performance in contrast to other methods, particularly improving the localization performance.

\begin{table}[t]
    \centering
    \caption{The performance of proposed methods on computationally synthesized scenes.}
    \begin{adjustbox}{width=\columnwidth,center}
    \begin{tabular}{c|c|cccc|c}
        \toprule[1pt]
         Method& Adapt. &$\mathrm{ER}_{20^{\circ}}\downarrow$ & $\mathrm{F}_{20^{\circ}}\uparrow$ &
         $\mathrm{LE}_\mathrm{CD}\downarrow$ & $\mathrm{LR}_\mathrm{CD}\uparrow$ & $\mathcal{E}_{\mathtt{SELD}}\downarrow$\\
         \midrule
         \multirow{2}{*}{SELD (Base + CSD)} &\XSolidBrush &0.739 &15.0\% &$30.0^\circ$ &30.8\% &0.612  \\
          &\CheckmarkBold &0.733 &15.4\% &$28.9^\circ$ &30.6\% &0.609  \\
         \midrule
         \multirow{2}{*}{Meta-SELD++} &\XSolidBrush &0.772 &13.9\% &$32.1^\circ$ &\textbf{32.0\%} &0.623  \\
          &\CheckmarkBold &0.745 &15.5\% &$30.4^\circ$ &31.8\% &0.610  \\
         \midrule
         \makecell[c]{Environment-Adaptive \\ Meta-SELD (Ours)} &\CheckmarkBold &\textbf{0.692} &\textbf{17.6\%} &$\mathbf{25.3^\circ}$ &31.3\% &\textbf{0.586}  \\
        \bottomrule[1pt]
    \end{tabular}
    \end{adjustbox}
    \label{tab: simulated scene}
\end{table}

\section{Conclusion}

This study presents environment-adaptive Meta-SELD designed for efficient adaptation to specific acoustic environments using a limited number of samples recorded in those settings. We apply Model-Agnostic Meta-Learning (MAML) to a pre-trained environment-independent SELD model to obtain generalized initial parameters for different environments. Subsequently, we introduce selective memory and environment representations in Meta-SELD++ to alleviate conflicts and the limitations of common initialization across different environments. When evaluated on the development set of the STARSS23 datasets and computationally synthesized scenes, our proposed environment-adaptive Meta-SELD demonstrates superior performance compared to conventional supervised-learning-based SELD methods. Furthermore, we investigate and exhibit environment representations. Experimental results show that environment representations effectively capture the nuances of diverse acoustic environments. The potential applications of these environment representations are extensive, promising significant advancements in enhancing acoustic scene analysis in diverse settings.

\appendix

As the microphones are mounted on an acoustically hard spherical baffle in the official setup of STARSS23, the frequency response of the $r$-th microphone with a wave number of $k$ on a rigid baffle of radius $R$ for $s$-th image source is \cite{archontisPhD}:
\begin{equation}
    H_{r s}\left(k, \psi_{r s}\right)=\sum_{n=0}^{\infty} \mathrm{i}^n(2 n+1) b_n(k R) P_n\left(\cos \psi_{r s}\right)
\end{equation}
where $\psi_{rs}$ denotes the angle between the DoA of the $s$-th sound source and the orientation of the $r$-th microphone, $P_n$ denotes the Legendre polynomial\cite{rafaely2015fundamentals, archontisPhD, politis2017comparing}, $\mathrm{i}$ is the imaginary unit, and $b_n$ is the mode strength term for a rigid baffle array given by \cite{rafaely2015fundamentals} 
\begin{equation}
    b_n(k R)=\frac{\mathrm{i}}{(k R)^2 h_n^{(1)^{\prime}}(k R)}
\end{equation}
with $h_n^{(1)^\prime}$ denoting the derivate of the $n$-th-order spherical Hankel function of the first kind\cite{rafaely2015fundamentals}.

The Ambisonics format conversion transforms the above-mentioned microphone-array format signals to first-order Ambisonics (FOA) format signals. The $n$-th-order and $m$-th-degree spherical harmonic function is defined with the angle $\Psi=\{\theta,\phi\}$ as follows \cite{rafaely2015fundamentals}:
\begin{equation}
Y_n^m(\Psi) \equiv \sqrt{\frac{2 n+1}{4 \pi} \frac{(n-m) !}{(n+m) !}} P_n^m(\cos \theta) e^{i m \phi}
\end{equation}
where $(\cdot)!$ represents the factorial operator, $\theta$ and $\phi$ are elevation and azimuth, and $P_n^m(\cdot)$ is the associated Legendre function. The spherical harmonic representation of the RIRs can be computed by using the following encoding process\cite{rafaely2015fundamentals, koyama2022spatial}:
\begin{equation}
    \mathbf{a}(k)=\mathbf{B}(k)^{-1} \mathbf{Y}^{\dagger} \mathbf{x}(k)
\end{equation}
where
\begin{equation}
\mathbf{B}(k)=\left(\begin{array}{cccc}
b_0 & 0 & 0 & 0\\
0 & b_1 & 0 & 0\\
0 & 0 & b_1 & 0\\
0 & 0 & 0 & b_1\\
\end{array}\right)
\end{equation}
$\mathbf{x}(k)$ denotes simulated microphone-array RIR signals, $(\cdot)^\dagger$ represents the Moore-Penrose pseudo inverse, and $\mathbf{a}(k)$ denotes the resulting FOA format signal. $\mathbf{Y}$ denotes first-order spherical harmony matrices with a four-channel microphone array as follows:
\begin{equation}
\mathbf{Y}=\left(
\begin{array}{cccc}
Y_0^0(\Psi_1) & Y_1^{-1}(\Psi_1) & Y_1^0(\Psi_1) & Y_1^1(\Psi_1) \\
Y_0^0(\Psi_2) & Y_1^{-1}(\Psi_2) & Y_1^0(\Psi_2) & Y_1^1(\Psi_2) \\
Y_0^0(\Psi_3) & Y_1^{-1}(\Psi_3) & Y_1^0(\Psi_3) & Y_1^1(\Psi_3) \\
Y_0^0(\Psi_4) & Y_1^{-1}(\Psi_4) & Y_1^0(\Psi_4) & Y_1^1(\Psi_4) \\
\end{array}\right)
\end{equation}
with $\Psi_r=\{\theta_r,\phi_r\}$ being the direction of the $r$-th microphone.

\scriptsize
\bibliographystyle{IEEEtran}
\bibliography{refs}

\end{document}